\documentclass[reprint,amsmath,amssymb, aps,
prb, floatfix]{revtex4-2}
\usepackage{ amssymb }

\usepackage{graphicx}
\usepackage{physics}
\usepackage{dcolumn}
\usepackage{color}
\usepackage{bm}
\usepackage{url} 
\usepackage{hyperref}
\hypersetup{
 colorlinks   = true,
 linkcolor=blue,
  citecolor    = blue,
  urlcolor=blue,
  } 
\usepackage[normalem]{ulem}
\usepackage[dvipsnames]{xcolor}
\newcommand{\vadim}{\textcolor{black}}
\newcommand{\outstrike}[1]{}

\begin{document}

\title{Slow electron-phonon relaxation controls the dynamics of the superconducting resistive transition}

\author{E.M. Baeva$^{1,2}$, A.I. Kolbatova$^{1,3}$, N.A. Titova$^{1}$, S. Saha$^{4}$, A. Boltasseva$^{4}$, S. Bogdanov$^{5,6,7}$, V.M. Shalaev$^{4}$, A.V. Semenov$^{1,8}$, A. Levchenko$^{9}$, G.N. Goltsman$^{2,10}$, and V.S. Khrapai$^{2,11}$}

\affiliation{$^1$ Moscow Pedagogical State University, Moscow 119435, Russian Federation \\ 
$^2$ National Research University Higher School of Economics, 20 Myasnitskaya Street, Moscow 101000, Russian Federation\\ 
$^3$ Laboratory of Photonic Gas Sensors, University of Science and Technology MISIS, Moscow 119049, Russian Federation \\
$^4$ Birck Nanotechnology Center and Elmore Family School of Electrical and Computer Engineering, Purdue University, West Lafayette, IN 47907, USA\\ 
$^5$ Department of Electrical and Computer Engineering, University of Illinois at Urbana-Champaign, Urbana, IL 61801, USA\\
$^6$ Holonyak Micro and Nanotechnology Lab, University of Illinois at Urbana-Champaign, Urbana, IL 61801, USA\\
$^7$ Illinois Quantum Information Science and Technology Center, University of Illinois at Urbana-Champaign, Urbana, IL 61801, USA\\
$^8$ Moscow Institute of Physics and Technology, Dolgoprudny, 141701, Russian Federation\\
$^9$ Department of Physics, University of Wisconsin-Madison, Madison, Wisconsin, WI 53706, USA
\\ $^{10}$ Russian Quantum Center, Moscow, Russian Federation \\ 
$^{11}$ Osipyan Institute of Solid State Physics, Russian Academy of Sciences, Chernogolovka, Russian Federation}

\begin{abstract}
We investigate the temporal and spatial scales of resistance fluctuations ($R$-fluctuations) at the superconducting resistive transition  accessed through voltage fluctuations measurements in thin epitaxial TiN films. This material is characterized by a slow electron-phonon relaxation, which puts it far beyond the applicability range of the textbook scenario of superconducting fluctuations. The measured Lorentzian spectrum of the $R$-fluctuations identifies their correlation time, which is nearly constant across the transition region and has no relation to the conventional Ginzburg-Landau time scale. Instead, the correlation time coincides with the energy relaxation time determined by a combination of the electron-phonon relaxation and the \vadim{relaxation via} diffusion \vadim{into} reservoirs. Our data is quantitatively consistent with the model of spontaneous temperature fluctuations and highlight \outstrike{important caveats in the accepted physical picture}\vadim{the lack of understanding} of the resistive transition in materials with slow electron-phonon relaxation.
\end{abstract}

\maketitle

\section{Introduction}

The phase transition of a normal conductor to the superconducting state can be imagined as a sudden drop of its resistance to zero at a critical temperature, $T_\mathrm{c}$, as in the very first measurement by Kamerlingh Onnes~\cite{KamerlinghOnnes1911}. In reality, the transition never occurs discontinuously in temperature ($T$) and its continuity is associated with spatial and temporal fluctuations of the modulus and phase of the superconducting order parameter. The imprint of these fluctuations on the DC resistance, $R$, or other time-averaged responses \vadim{is easily observable in samples of reduced dimensionality and} can be revealed by zooming into the transition region with a variety of fluctuation spectroscopy approaches~\cite{Varlamov2018}. Shifting the focus\outstrike{ from static responses} towards the \outstrike{dynamics}\vadim{dynamic responses, including time-domain~\cite{Jelic2016-ro,Spahr2020,Mannila2021}, frequency-domain~\cite{Karasik1996,Burke1998,Kardakova2013} or noise response~\cite{VanOoijen1965,McCumberHalperin,Clem1981,Golubev2008},} helps to reveal microscopic origin and intrinsic time-scales of the superconducting fluctuations.

The textbook~\cite{Larkinlate2005} scenario of the resistive transition (RT) is based on the time-dependent Ginzburg-Landau (GL) equation that defines a universal time-scale of the superconducting fluctuations -- the GL-time -- $\tau_\mathrm{GL}=\pi \hbar/8 k_\mathrm{B} (T-T_\mathrm{c})$. Here and below we concentrate on the \vadim{narrow temperature range of the strongest $R(T)$ dependence}\outstrike{region above $T_\mathrm{c}$}, where global superconducting state is not yet set in and the physics of the Berezinskii-Kosterlitz-Thouless (BKT) phase transition~\cite{Halperin1979,Benfatto2009} and  thermal~\cite{LangerAmbegaokar,McCumberHalperin,Golubev2008,Xavier2016} and quantum~\cite{Konig2021,Bezryadin2000, GolubevZaikin2001,Lau2001,Arutyunov2008,Bae2009} phase-slips is irrelevant. Accordingly,  the $\tau_\mathrm{GL}$ defines the correlation time and correlation length $\xi_\mathrm{GL} = \sqrt{\mathcal{D}\tau_\mathrm{GL}}$ of the superconducting order parameter fluctuations which underpin the  Aslamazov-Larkin (AL)~\cite{Aslamasov1968,Breznay2012} and the Maki-Thompson~\cite{Maki1968,Thompson1970} paraconductivity contributions. This appealingly simple picture \vadim{constitutes the commonly accepted scenario of the RT, which} is valid under the assumption of sufficiently fast electron-phonon (e-ph) relaxation, characterized by the relaxation time $\tau_\mathrm{eph}$, namely for $\tau_\mathrm{eph}\ll \tau_\mathrm{GL}$. In the AL theory, this allows one to neglect so-called nonlinear fluctuation effects, which increase the correlation time and make the AL contribution more singular at $T_\mathrm{c}$~\cite{Larkin2001,Larkinlate2005}. More generally, it is this assumption, also known as the approximation of local equilibrium~\cite{Kramer1978,WattsTobin1981}, that underpins the microscopic derivation of the whole set of dynamic equations of superconductivity, including the time-dependent GL equation~\cite{Kramer1978,WattsTobin1981,Golub1976,Schoen1979,Ivlev1984}. In local equilibrium spontaneous fluctuations associated with the fast energy exchange between the electrons and acoustic phonons effectively average out during a much slower fluctuation of the superconducting order parameter.
 
Over the last \vadim{approximately six} decades, numerous experimental and theoretical works have addressed the problem of fluctuations at the RT in thin superconducting films~\cite{VanOoijen1965,Wessels2021,Gottardi2021,Petkovic2013,Bagrets2014,Zhang2008,Reznikov2006,Luukanen2003,Hoevers2000,Burke1999,Ekstrom1995,Nagaev1991, Knoedler1983, Knoedler1982,Voss1980,Weitzel_PRL_2023}, without reaching a \outstrike{consensus about }\vadim{full understanding of} their microscopic origin and  time-scales. To our best knowledge, no  experimental evidence of the \vadim{relevance of the} GL-time at the RT exists and this may not be surprising, since the applicability range of the time-dependent GL theory in terms of the ratio $\tau_\mathrm{eph}/ \tau_\mathrm{GL}$ is usually too narrow~\cite{WattsTobin1981}. Thus, the microscopic origin of the dynamics of the superconducting fluctuations at the RT, especially in the limit of large $\tau_\mathrm{eph}$, remains an open question of fundamental importance.

\section{Experimental idea and samples}

In this manuscript, we investigate the \outstrike{dynamics of }\vadim{noise response near} the RT in thin films of TiN of epitaxial quality. This material is characterized by \outstrike{a record }slow e-ph relaxation, $\tau_\mathrm{eph}\sim 5\,\mathrm{ns}\gg \tau_\mathrm{GL}$ at $T_\mathrm{c}\approx4\,$K, and falls beyond the applicability range of the \outstrike{textbook scenarios }\vadim{commonly accepted scenario}. \vadim{Here and below the $T_\mathrm{c}$ is taken as the midpoint of the RT, which is legitimate for quasi-one dimensional samples studied in this work.} The \vadim{measured} voltage \outstrike{drop on the device }is characterized as a stationary random process that takes its origin in spontaneous fluctuations of the resistance ($R$-fluctuations). We \outstrike{measure }\vadim{investigate} the noise spectra within a 100~MHz bandwidth and obtain both the variance and the correlation time $\tau_\mathrm{R}$ of the $R$-fluctuations using the Wiener-Khinchin theorem~\cite{kogan2008electronic}. Near the steepest part of $R(T)$ curve, $\tau_\mathrm{R}$ is nearly temperature independent and coincides with the thermal relaxation time, mediated by a competition between the e-ph relaxation and diffusion \vadim{into} the reservoirs. Our results unequivocally identify spontaneous fluctuations of temperature ($T$-fluctuations) as the origin of the $R$-fluctuations and demonstrate \vadim{that} the \outstrike{textbook }\vadim{accepted} fluctuation scenario of the RT is incomplete. 

Our samples are \outstrike{microbridges }patterned from epitaxial TiN films of 5\,nm to 20\,nm thickness ($d$) grown on (111) sapphire substrates using DC reactive magnetron sputtering~\cite{Saveskul}. The fabrication and transport characterization are detailed in the Supplemental Material (SM)~\cite{Suppl_Data} and in \vadim{Ref.}~\cite{Second_manuscript}, respectively. In contrast to previously studied strongly disordered TiN films~\cite{Sacepe2008, Sacepe2010, Gao2012, Driessen2012,Kardakova2013, Postolova2017}, the present material exhibits a uniform monocrystalline structure supported by the X-ray analysis and atomic force microscopy~\cite{Naik2012,Kinsey14,Saveskul}. Consistently, recent transport studies revealed exceptional electronic properties, including surface scattering-limited mean free path, close to the bulk value of $T_\mathrm{c}$, and \vadim{moderate}\outstrike{ minute} surface magnetic disorder~\cite{Saveskul}. 

\begin{figure}[t!]
    \includegraphics[scale=1]{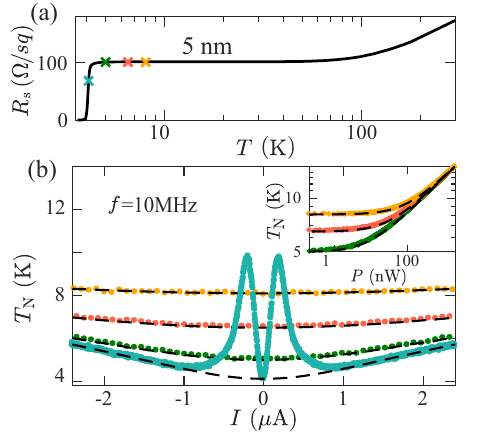}
    \caption{\label{figure_n_2} Noise thermometry along the $R(T)$ curve. (a) The $T$ dependence of sheet resistance $R_\mathrm{s}$ for sample A1. (b) Main: current dependencies of $T_\mathrm{N}$ for sample A1 at $T_\mathrm{b}$ marked in (a) with the same color. Inset: $T_\mathrm{N}(P)$ dependencies at 5\,K, 6.5\,K, and 8\,K on a log-log scale.}
\end{figure}

\section{Noise thermometry in the normal state}

The noise measurements were performed on five devices patterned in the form of two-terminal rectangular bridges. The characteristics of all studied samples, including their dimensions, are given in Table I in the SM~\cite{Suppl_Data}. Here we mainly focus on the data from two $d=5$\,nm devices for which the length and width of the bridges are $L\approx9.7\,\mathrm{\mu m}$, $w\approx450\,\mathrm{nm}$ (A1) and $L\approx2.8\,\mathrm{\mu m}$, $w\approx110\,\mathrm{nm}$ (A2). In both devices a textbook-like $R(T)$ dependence is found and displayed in \autoref{figure_n_2}(a) for device A1. \vadim{Namely, at} decreasing $T$ \vadim{from about 50\,K }the residual resistance plateau is seen followed by a sharp transition to the superconducting state. The symbols mark discrete values of $T$ where the noise temperature, $T_\mathrm{N}$, was measured as a function of the DC bias current $I$. Here, $T_\mathrm{N}$ reflects the spectral density of the voltage fluctuations 
\vadim{via a Johnson-Nyquist relation~\cite{kogan2008electronic}. The voltage fluctuations are measured} 
within a narrow frequency band ($<1$\,MHz) centered at the resonance of the tank circuit ($\approx10$\,MHz). Details of the noise measurements are discussed in the SM~\cite{Suppl_Data}. \autoref{figure_n_2}(b) compares the data in the normal state and at the RT. In the former case (three upper curves), the observed gradual increase of $T_\mathrm{N}$ at increasing $I$ is determined by the e-ph cooling. As shown in the inset, in a wider range of bias currents the data is well described by a dependence $P=\mathcal{V}\Sigma_\mathrm{eph}(T_\mathrm{N}^5-T^5)$, where $P$ is the Joule heating power, $\mathcal{V}$ is the volume of the device and $\Sigma_\mathrm{eph}=1.15\cdot 10^8\,\text{WK}^{-5}\text{m}^{-3}$ is the e-ph cooling power, see the dashed lines fits (the same in the inset and in the body of \autoref{figure_n_2}(b)). From this data we determine the e-ph relaxation length via $L_\mathrm{eph} = \sqrt{\sigma\mathcal{L}/5\Sigma_\mathrm{eph}T^3}$, where  $\sigma$ is the conductivity and $\mathcal{L}$ is the Lorenz  number~\cite{Denisov2020}, which corresponds to $L_\mathrm{eph} =1.2\pm0.1\,\mu\mathrm{m}$ at $T=T_\mathrm{c}$. By contrast, the curve taken at the RT exhibits giant noise at small currents and meets the prediction for the e-ph cooling only at sufficiently high $\abs{I}$. In the following, we argue that this giant noise comes from the $R$-fluctuations with the time-scale given by the relaxation to the thermal bath. 

\section{Resistance fluctuations at the superconducting transition}

\begin{figure}[t]
    \includegraphics[scale=1]{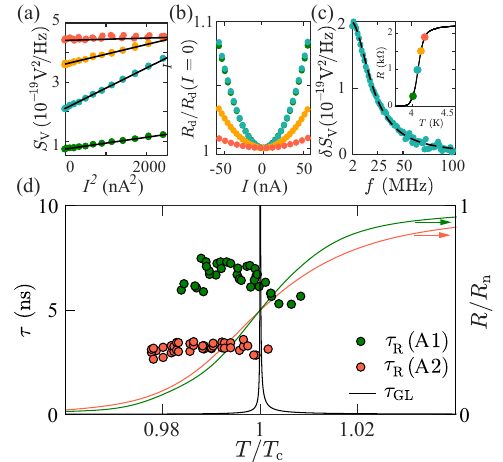}
    \caption{\label{figure_n_3} Spectroscopy of the $R$-fluctuations. Measured $S_\mathrm{V}$ vs $I^2$ (a) and the normalized $R_\mathrm{d}$ vs $I$ (b) for sample A1 measured at $T_\mathrm{b}$ marked by the same symbols in RT (inset of (c)). (c) The typical frequency dependence of normalized $\delta S_\mathrm{V}$ is shown with symbols on a linear scale. (d) The correlation time $\tau_\mathrm{R}$ and normalized resistances for samples A1 and A2 vs $T/T_\mathrm{c}$.}
\end{figure}

\autoref{figure_n_3}(a) demonstrates that at small currents the noise obeys a parabolic dependence on $I$, which is a clear signature of spontaneous $R$-fluctuations~\cite{kogan2008electronic}. The spectral density of the voltage fluctuations equals $S_\mathrm{V} = 4k_\mathrm{B}TR_\mathrm{d}+\delta S_\mathrm{V}$, where the first term is the Johnson-Nyquist noise\vadim{, evaluated with the help of the differential resistance $R_\mathrm{d}\equiv \dd V/\dd I$}. The excess noise is given by $\delta S_\mathrm{V}= I^2 S_\mathrm{R}$, where $S_\mathrm{R}$ is the spectral density of the $R$-fluctuations. $\delta S_\mathrm{V}$ depends on the chosen $T$ and is maximum near the \outstrike{center }\vadim{midpoint} of the RT, see the locations on the $R(T)$ curve marked by corresponding symbols in the inset of \autoref{figure_n_3}(c). \vadim{A correspondence between the $R(T)$ curve and the dependencies $S_\mathrm{V}(T)$ at different fixed $I$ in a $d=20$\,nm device is given in the SM (Fig.~S4)}. As shown in \autoref{figure_n_3}(b), within the same range of $I$ the $R_\mathrm{d}$ changes at most by 10\%, meaning that $\delta S_\mathrm{V}$ essentially captures the equilibrium $R$-fluctuations.

The measurement of the correlation time is achieved by removing the tank circuit at the input of the low-$T$ amplifier and measuring the spectrum of the voltage fluctuations within a 100 MHz bandwidth. \vadim{This type of measurement is performed after stabilization of the slowly drifting bath temperature to within $\pm2.5$\,mK.} The calibration of the frequency-dependent gain is a technically challenging procedure, detailed in the SM~\cite{Suppl_Data}. As shown in \autoref{figure_n_3}(c), we observe a Lorentzian spectrum $\delta S_\mathrm{V}(f)/\delta S_\mathrm{V}(0) = (1+\omega^2\tau_\mathrm{R}^2)^{-1}$, where $\omega = 2\pi f$, see the dashed line fit. According to the Wiener-Khinchin theorem, \vadim{the spectral density $S_\mathrm{R}\propto\delta S_\mathrm{V}$ equals twice the Fourier transform of the time correlation function of the $R$-fluctuations. Taking the inverse Fourier transform we get:}\outstrike{this evidences the correlation function:}
\begin{equation}
\label{eq1}
	\langle\delta R(t)\delta R(0) \rangle = \langle\delta R^2 \rangle \exp(-t/\tau_\mathrm{R}),
\end{equation}
where $\langle\delta R^2 \rangle$ and $\tau_\mathrm{R}$ are, respectively, the variance and the correlation time of the R-fluctuations. 

In \autoref{figure_n_3}(d) we plot $\tau_\mathrm{R}$ obtained at different points across the RT in both devices. Also shown are the normalized $R(T)$ curves as a function of $T/T_{c}$. We find that $\tau_\mathrm{R}$ remains approximately constant around the steepest part of the transition, where \outstrike{the voltage fluctuations are maximum and }$R(T)$ changes by almost an order of magnitude within a temperature range of $\sim 100\,$mK. This is in sharp contrast with the divergent behavior expected for the GL time, shown by the solid line. Note that $\tau_\mathrm{R}$ lies in the range of a few nanoseconds by far exceeding $\tau_\mathrm{GL}$ everywhere except for the immediate vicinity of this divergence. Notably, $\tau_\mathrm{R}$ depends on the dimensions of the device being roughly twice smaller in a shorter device A2. The origin of this size effect becomes clear from the $T$-dependence of $\tau_\mathrm{R}$ in a much wider temperature range, as explained below.

\begin{figure}[t]
    \includegraphics[scale=1]{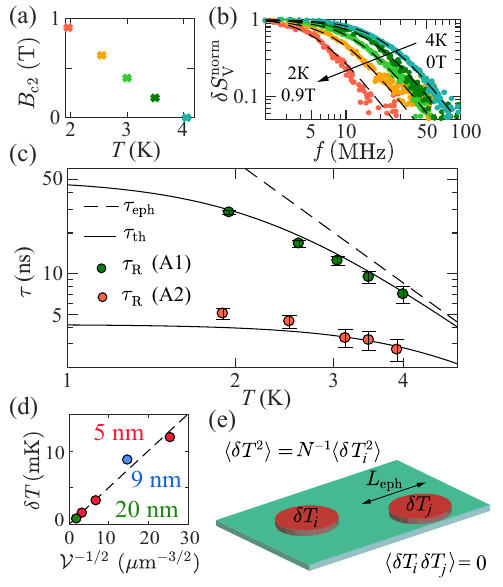}
    \caption{\label{figure_n_4} Evidence of the $T$-fluctuation scenario. The $T$ dependence of $B_\mathrm{c2}$ (a) and normalized spectra of $\delta S_\mathrm{V}$ on a log-log scale (b) for sample A1. The black dashed lines represent Lorentzian fits used to obtain $\tau_\mathrm{R}$. The spectra are measured at different bath temperatures $T$ using the perpendicular magnetic field $B_\mathrm{c2}$. (c) The $T$ dependencies of $\tau_\mathrm{R}$ for samples A1 and A2 in comparison with $\tau_\mathrm{th}$ and $\tau_\mathrm{eph}$. (d) $T$-fluctuation as a function of sample volume $\mathcal{V}$. (e) Spatial distribution of the $T$-fluctuations with correlation length given by $L_\mathrm{eph}$. Partial fluctuations $\delta T_i$ and $\delta T_j$ at distances exceeding $L_\mathrm{eph}$ are independent.}
\end{figure}

We extend the $T$-range of the fluctuation measurements by shifting $T_\mathrm{c}$ in a magnetic field, $B$, applied perpendicular to the film. The $T$-dependence of the second critical magnetic field, $B_\mathrm{c2}$, that determines the $B$-$T_\mathrm{c}$ correspondence is displayed in \autoref{figure_n_4}(a). The normalized $\delta S_\mathrm{V}(f)$ spectra obtained in each of these points are shown in \autoref{figure_n_4}(b) for device A1. At decreasing $T_\mathrm{c}$, the Lorentzian spectra span progressively narrower frequency range, signaling the increase of $\tau_\mathrm{R}$, see the dashed line fits. The measured $T$-dependencies are plotted in \autoref{figure_n_4}(c) for both devices (symbols). In the longer device A1, the absolute value of $\tau_\mathrm{R}$ at higher $T$ is close to the e-ph relaxation time $\tau_\mathrm{eph}\equiv L_\mathrm{eph}^2/\mathcal{D}$ obtained from the cooling rate $\Sigma_\mathrm{eph}$ in the normal state and diffusion coefficient $\mathcal{D}$, see the dashed line. However, the $T$-dependence of the $\tau_\mathrm{R}$  is considerably weaker than $\tau_\mathrm{eph}\propto T^{-3}$. In the shorter device A2 the $T$-dependence of the $\tau_\mathrm{R}$ is even weaker and indicates a saturation in the low-$T$ limit. We explain this data by identifying $\tau_\mathrm{R}$ with the thermal relaxation time that takes into account a combined effect of the e-ph relaxation and \vadim{relaxation via} diffusion \outstrike{in }\vadim{into} the source and drain reservoirs: $\tau_\mathrm{th} = (\tau_\mathrm{eph}^{-1}+\tau_\mathcal{D}^{-1})^{-1}$~\cite{Karasik1996,Burke1998,reulet2005} with $\tau_\mathrm{eph}=L_\mathrm{eph}^2/\mathcal{D}$ and $\tau_\mathcal{D} = \left(L/\pi\right)^2/\mathcal{D}$, where $L_\mathrm{eph}$ is the e-ph relaxation length and $\mathcal{D}$ is the diffusion coefficient. $\tau_\mathrm{th}$ is shown by solid lines in \autoref{figure_n_4}(c) obtained with $\mathcal{D} = 1.8$\,cm$^2$/s, the value consistent with an independent estimate $\mathcal{D}\approx2.5$\,cm$^2$/s from the slope of $\dd B_\mathrm{c2}/\dd T$. Note that $\tau_\mathcal{D}$ is different from the usual dwell time used in the analysis of hot-electron bolometers~\cite{Karasik1996,Burke1999} and represents the slowest among the set of the relaxation times of the diffusion relaxation, see the derivation in the SM~\cite{Suppl_Data}. The agreement between $\tau_\mathrm{R}$ and $\tau_\mathrm{th}$ explains that the data in \autoref{figure_n_3}(d) manifests the size effect in thermal relaxation~\cite{extrinsic_remark} which occurs at $L\approx\pi L_\mathrm{eph}$.

\section{Spontenaous $T$ fluctuations}

\autoref{figure_n_4}(c) points to stochastic interaction with the thermal bath\vadim{, represented by both the phonon bath and the electronic reservoirs,} as the origin of the $R$-fluctuations. Absorption/emission of individual acoustic phonons and diffusion of individual carriers \vadim{into}/out of the reservoirs both generate random energy exchange with the bath. In thermodynamics, corresponding fluctuations of the free-energy of the electron system within a given volume $\mathcal{V}$ are expressed via spontaneous $T$-fluctuations: $\langle\delta F^2 \rangle = (C_\mathrm{e}\mathcal{V})^2\langle\delta T^2 \rangle$, where $C_\mathrm{e}$ is the electronic heat capacity. Therefore, the textbook expression for the variance of the $T$-fluctuations~\cite{LandauLifshitz1980} $\delta T^2\equiv\langle\delta T^2 \rangle = k_\mathrm{B}T^2/C_\mathrm{e}\mathcal{V}$ has a transparent interpretation -- $\delta T$ is a measure of how close the system can approach the superconducting phase in the course of thermal fluctuations. The expression for $\delta T$ is sometimes applied to estimate the $R$-fluctuations from the slope of $R(T)$ dependence as $\langle\delta R^2 \rangle = \abs{\dd R/\dd T}^2\delta T^2$~\cite{Voss1976,Kogan1985, Ekstrom1995, Hoevers2000}. Our experiment allows to directly verify this expression based on the measured noise spectra and the identity $\langle\delta R^2 \rangle = \int S_\mathrm{R} \dd f = S_\mathrm{R}(0)/4\tau_\mathrm{R}$. This is achieved for a fixed $T$ near the mid-point of the RT, where both the $\dd R/\dd T$ and the $S_\mathrm{R}$ are maximum (see the SM~\cite{Suppl_Data} for the correspondence between these two quantities). \autoref{figure_n_4}(d) summarizes the data on RMS fluctuation $\delta T$ obtained in all studied devices. We find indeed that $\delta T\propto \mathcal{V}^{-1/2}$ and the data is quantitatively consistent with the above thermodynamic expression (dashed line). Thus, in contrast to giant low-frequency noises mediated by flux-flow~\cite{Clem1981} and fluctuations in percolating networks~\cite{Kiss_PRL_1993}, our experiment reveals a more universal scenario of a broadband noise at the RT~\cite{Hoevers2000,Zhang2008,Voss1980, Knoedler1982, Knoedler1983,Ekstrom1995, Burke1999,Maul1969,VanOoijen1965}.

How local are the $T$-fluctuations? The scaling ${S_\mathrm{R}\propto\delta T^2 \propto\mathcal{V}^{-1}}$ implies spatial averaging of the $R$-fluctuations and the underlying $T$-fluctuations, which can be understood by considering their correlation length $L_\mathrm{eph}$. Fluctuations separated by a larger distance belong to different correlation volumes, and they are statistically independent, see the sketch of \autoref{figure_n_4}(e). In the experimentally relevant 1D case ($w < L_\mathrm{eph}$) one can estimate the fluctuation of the resistance $R_i$ in each  correlation volume as $\langle\delta R_i^2 \rangle = \langle\delta R^2 \rangle/N$, where $i\in [1,N]$, $N = L/L_\mathrm{eph}$ and $\langle R_i \rangle = R/N$. In our different samples, the normalized RMS fluctuation $\sqrt{\langle\delta R_i^2 \rangle}/(R_\mathrm{n}/N)$, where $R_\mathrm{n}$ is the normal state resistance, reaches from 0.1 to 0.25 near the maximum of $\dd R/\dd T$. Note that the absolute possible maximum equals 1/2, and it is achieved when $R_i$ switches between 0 and $R_\mathrm{n}/N$ in a random telegraph noise process, see the SM~\cite{Suppl_Data}. This demonstrates that in spite of vanishing in the limit of $\mathcal{V}\rightarrow\infty$ the $T$-fluctuations remain important, and they can locally drive the system a sizeable fraction of the way between the normal and the superconducting state.

\section{Discussion: slow vs fast relaxation}

As seen from \autoref{figure_n_3}(d), our devices fall in the regime of slow e-ph relaxation $\tau_\mathrm{eph}\sim\tau_\mathrm{R}\gg\tau_\mathrm{GL}$, a situation opposite to the \outstrike{textbook }\vadim{standard} case of fast relaxation. In this case, much faster superconducting fluctuations have enough time to adopt to a local $T$-fluctuation $\delta T_i$. In principle, this may lead to extra broadening of the RT, provided $\sqrt{\langle\delta T_i^2\rangle} > Gi$, where $Gi$ is the Ginzburg-Levanyuk number that controls the transition width in the AL theory. The analysis of this hypothesis lies beyond the scope of the present work and is performed in a \vadim{separate paper}~\cite{Second_manuscript}. Here it is illuminating to discuss the fate of the $T$-fluctuations in a conventional regime of $\tau_\mathrm{eph}\ll\tau_\mathrm{GL}$. In this regime, the $T$-fluctuations average both in space, over the GL volume $\mathcal{V}_\mathrm{GL}$, and in time, during the GL-time. The average fluctuation is therefore given by $\langle\delta T^2\rangle_\mathrm{av} = (k_\mathrm{B}T^2/C_\mathrm{e}\mathcal{V}_\mathrm{GL})\times(4\tau_\mathrm{eph}/\tau_\mathrm{GL})$. The last factor takes into account that only sufficiently low-frequency components survive temporal averaging so that for the infinitely fast e-ph relaxation ($\tau_\mathrm{eph}\rightarrow0$) the $T$-fluctuations are irrelevant. More rigorously, one can neglect them given $\sqrt{\langle\delta T^2\rangle_\mathrm{av}} < (T-T_\mathrm{c})$. As derived in the SM~\cite{Suppl_Data}, this requires $\tau_\mathrm{eph} \ll \hbar/(k_\mathrm{B}T_\mathrm{c}Gi)$, that is in our case $\tau_\mathrm{eph} \ll 1$\,ns (a more optimistic estimate for the device A1). The fact that the actual e-ph relaxation is much slower explains the relevance of the $T$-fluctuations in the present experiment and the breakdown of the \outstrike{textbook }\vadim{accepted} scenario of the dynamics of the RT. \vadim{Note that the above condition to neglect the $T$-fluctuations is rather strong and may not be fulfilled also in Al~\cite{Ivlev1984}, Nb~\cite{Gershenzon1984} and other superconductors with not too fast e-ph relaxation. Whether the extra broadening of the RT owing to the $T$-fluctuations is directly observable in such materials is a matter of future experiments. }

In summary, we have shown that the fluctuation dynamics at the RT in \outstrike{high-quality }\vadim{epitaxial} thin TiN films is governed by the $T$-fluctuations mediated by stochastic energy exchange with the thermal bath. These fluctuations have a well-defined correlation time and correlation length, which coincide with the corresponding scales of the energy relaxation. The fundamental reason for this behavior is the extremely slow e-ph relaxation as compared to the GL time, $\tau_\mathrm{eph}\gg\tau_\mathrm{GL}$, that puts this system beoynd the applicability range of the \outstrike{textbook fluctuation scenario }\vadim{accepted scenario of superconducting fluctuations}. Our findings reveal \outstrike{caveats in the accepted }\vadim{that the present} understanding of the RT \vadim{is incomplete} and indicate the promising direction for future research.

\section*{Acknowledgments}

We are grateful to I. Burmistrov, A. Denisov, M. Feigelman, I. Gornyi, F. Jaekel, E. K\"{o}nig, D. McCammon, R. McDermott, A. Melnikov, D. Shovkun, A. Shuvaev and E. Tikhonov for fruitful discussions. This study was conducted as a part of strategic project “Digital
Transformation: Technologies, Effectiveness, Efficiency” of Higher School of Economics development programme granted by Ministry of science and higher education of Russia “Priority-2030” grant as a part of “Science and Universities” national project and the Basic Research
Program of the HSE University (transport measurements). The work was partially funded by the Ministry of Science and Higher Education of the Russian Federation FSME-2022-0008 (chip fabrication). The work at the UW-Madison was supported by the U.S. Department of Energy (DOE), Office of Science, Basic Energy Sciences (BES) under Award No. DESC0020313 (A.L.). 

\nocite{Caprara2011,Kibble_2007,Kamenev,Levchenko2007,Kwak-AP2023,Levchenko2010,Irwin2005,blanter2000shot,Huard2007,Diroll2020,Kittel,Swartz1989,Kim1992,tikhonov2013equations}


\begin{thebibliography}{82}
\bibitem [{ext()}]{extrinsic_remark}%
  \BibitemOpen
  \href@noop {} {}\bibinfo {note} {As a side remark, we note that the observed
  correspondence of $\tau_\mathrm{R}$ with the microscopic timescales allows to
  exclude possible extrinsic origin of the measured wide band voltage noise,
  as, e.g., contact noise, pick-up or fluctuations of the bath
  temperature.}\BibitemShut {Stop}%
\end{thebibliography}

%

\end{document}


\title{SUPPLEMENTAL MATERIAL for \\ Slow electron-phonon relaxation controls the dynamics of the superconducting resistive transition}

\author{E.M. Baeva$^{1,2}$, A.I. Kolbatova$^{1}$, N.A. Titova$^{1}$, S. Saha$^{3}$, A. Boltasseva$^{3}$, S. Bogdanov$^{4,5,6}$, V.M. Shalaev$^{3}$, A.V. Semenov$^{2}$, A. Levchenko$^{7}$, G.N. Goltsman$^{2,8}$, and V.S. Khrapai$^{2,9}$}

\affiliation{$^1$Moscow Pedagogical State University, Moscow, Russia\\$^2$HSE University, Moscow, Russia\\$^3$Birck Nanotechnology Center and Elmore Family School of Electrical and Computer Engineering, Purdue University, West Lafayette, IN 47907, USA\\$^4$Department of Electrical and Computer Engineering, University of Illinois at Urbana-Champaign, Urbana, IL 61801, USA\\$^5$Holonyak Micro and Nanotechnology Lab, University of Illinois at Urbana-Champaign, Urbana, IL 61801, USA\\$^6$Illinois Quantum Information Science and Technology Center, University of Illinois at Urbana-Champaign, Urbana, IL 61801, USA\\$^7$Department of Physics, University of Wisconsin-Madison, Madison, Wisconsin, WI 53706, USA\\$^8$Russian Quantum Center, Moscow,
Russia\\$^9$Osipyan Institute of Solid State Physics, Russian Academy of Sciences, Chernogolovka, Russia}

\begin{abstract}
In this supplemental material we (i) derive from the first principles an upper estimate of the spectral density of the resistance fluctuations, its average over the sample volume and its dependence on the correlation time; (ii) present analysis for a novel mechanism of intrinsic excess noise, that originates from the correlated fluctuations of conductivity in the regime when electronic energy relaxation is slow compared to the time scale of dynamic superconducting fluctuations; (iii) provide numerical estimates for the strength of this noise as compared to the empirical $T$-fluctuations mechanism discussed in the main text; \anna{(iv) provide experimental details including description of sample parameters, setup for the noise thermometry, results of the noise thermometry far above $T_\mathrm{c}$ and along the resistive transition, setup for the noise spectroscopy; (v) estimate the Kapitza resistance and (vi) derive the energy relaxation time due to electron diffusion into contacts.}  
\end{abstract}


\maketitle


\section{I. Estimate of the resistance noise}

\subsection{Starting point}

In the following we provide an estimate of the spectral density of the resistance fluctuations ($R$-fluctuations), from the first principles. We assume a specific form of temporal correlation function of the $R$-fluctuations:
\begin{equation}
	\ev{\delta R(t)\delta R(0)} = \langle \delta R^2\rangle \exp(-t/\tau_\mathrm{R}), \label{eq:corrfunc}
\end{equation}

where $\tau_\mathrm{R}$ is the correlation time of the $R$-fluctuations and $\langle \delta R^2\rangle$ is their variance. At the superconducting transition the instantaneous value obeys $0\leq R(t)\leq R_\text{n}$, where $R_\text{n}$ is the normal-state value of the resistance. It is straightforward to show that for a given average resistance $\langle R\rangle$ the variance is bounded by:

\begin{equation}
	\langle \delta R^2\rangle\leq \langle R\rangle\left(R_\text{n}-\langle R\rangle\right),  \label{eq:max}
\end{equation}
with the maximum realized in the case of $R$-fluctuation between the values of $R=0$ and $R=R_\text{n}$ (i.e. the simplest case of a telegraph noise). This observation directly follows from the observation that for a random quantity fluctuating in the range $0<x<1$ with the fixed average value $\langle x\rangle$, at any instance $x^2\leq x$, hence $\langle \delta x^2\rangle\leq\langle x\rangle\left(1-\langle x\rangle\right)$, and the equality take place only for bimodal fluctuations between 0 and 1. Using the  Eqs.~(\ref{eq:corrfunc},\ref{eq:max}) one obtains from the Wiener-Khinchin theorem the upper bound on the spectral density of the $R$-fluctuations at a given frequency $\omega = 2\pi f$:
\begin{equation}
	S_\mathrm{R} (\omega) \leq \langle R\rangle\left(R_\text{n}-\langle R\rangle\right)\frac{4\tau_\mathrm{R}}{1+\omega^2\tau_\mathrm{R}^2}.  \label{eq:spectral}
\end{equation}

Now we can calculate the $R$-fluctuations in a large sample of a thin superconducting film, that is the sample with area $A$ much larger than the correlation area of the fluctuations $A_\text{c}$. In two-dimensions (2D) $A_\text{c} = \mathcal{D}\tau_\mathrm{R}$ and in one dimension (1D) $A_\text{c} = w\sqrt{\mathcal{D}\tau_\mathrm{R}}$, where $w<\sqrt{\mathcal{D}\tau_\mathrm{R}}$ is the width of the sample and $\mathcal{D}$ is the diffusion coefficient. The total $R$-fluctuations of the sample are determined via averaging of the fluctuations within individual correlation areas, which are by definition independent. 

\subsection{Calculation in 1D} 
The derivation in 1D is straightforward, since the total resistance is just a sum of the resistances $R =\sum R_i$, where $i\in [1:N=A/A_\text{c}]$. Having in mind that $R_i$ fluctuate independently and using Eq.~\eqref{eq:spectral} we get:
\begin{equation*}
	S_\mathrm{R} = \sum_1^N S_{\mathrm{R}_i}\leq \frac{\langle R\rangle\left(R_\text{n}-\langle R\rangle\right)}{N}\frac{4\tau_\mathrm{R}}{1+\omega^2\tau_\mathrm{R}^2},  
\end{equation*}
and, finally, in the low-frequency limit $\omega\rightarrow0$:
\begin{equation}
	S_\mathrm{R} (\omega=0)\leq 4\tau_\mathrm{R}\frac{\langle R\rangle\left(R_\text{n}-\langle R\rangle\right)}{N} = 4\frac{1}{L}\langle R\rangle\left(R_\text{n}-\langle R\rangle\right)\mathcal{D}^{1/2}\tau_\mathrm{R}^{3/2}, \label{eq:1D}
\end{equation}
where $L$ is the length of the 1D wire.

\subsection{Calculation in 2D}

In 2D the resistance is independent of the size that slightly complicates the derivation. The main obstacle is the need to consider the percolation problem, that has been widely discussed in the context of $1/f$ noise in metal-insulator and metal-superconductor inhomogeneous networks, see S.M. Kogan's book for a review~\cite{kogan2008electronic}. For simplicity, we will not consider a percolation problem here. Instead, we will approach the problem with the help of the analytical model of the effective medium theory (EMT), developed in Refs.~\cite{Benfatto2009,Caprara2011}. 

For the purpose of estimate of the $R$-fluctuations we assume that the 2D (square) film of the area $A$ consists of $N=A/A_\mathrm{c}$ independently fluctuating (square) regions. In the normal state the resistance of the film, as well as the resistance of each of such regions is given by $R_\mathrm{n}$. At the resistive transition the fraction of the normal regions, denoted as $p$, decreases below 1. Again, the maximum possible fluctuation is achieved when the resistance of each region fluctuates between 0 and $R_\mathrm{n}$. In this case, the resistance of the film according to the EMT~\cite{Benfatto2009,Caprara2011} is simply given by:
\begin{equation}
	 R = 2R_\mathrm{n}\left(p-\frac{1}{2}\right) \label{eq:R_EMT}  
\end{equation}

In the EMT, Eq.~\eqref{eq:R_EMT} is used to expresses the average resistance $\langle R \rangle$ via the average fraction $\langle p \rangle$ of the normal regions. Here we will use the same relation to calculate the resistance noise, mediated by the noise of $p$, very similar to the ideas used to explain $1/f$ noise in high-$T_\mathrm{c}$ compounds by Kiss and Svedlindh in Ref.~\cite{Kiss_PRL_1993}. Under the assumption of independent fluctuations in each of $N$ regions, we obtain for the fluctuation spectral density of the fluctuations of $p$: 
\begin{equation*}
	 S_\mathrm{p} = \frac{4\tau_\mathrm{R}}{1+\omega^2\tau_\mathrm{R}^2}\cdot \frac{p\left(1-p\right)}{N},
\end{equation*}
Thus, using Eq.~\eqref{eq:R_EMT}, we obtain the upper limit for the $R$-fluctuations of the whole sample at $\omega =0$:

\begin{equation}
	 S_\mathrm{R}(\omega =0)  \leq 4R_\text{n}^2S_\text{p}(\omega =0) = \frac{4\mathcal{D}\tau_\mathrm{R}^2}{A}\left(R_\text{n}^2 - \langle R\rangle^2\right), \label{eq:2D}
\end{equation}

where we have used that $A_\mathrm{c} = \mathcal{D}\tau_\mathrm{R}$ in the 2D case. Note a stark difference of this result from the 1D estimate of Eq.~\eqref{eq:1D}. In the 2D case, the maximum $R$-fluctuation corresponds to the lowest point of the resistive transition  $\langle R\rangle = 0$, whereas in the 1D case the maximum is in the center of the transition $\langle R\rangle = R_\text{n}/2$. Interestingly, in the framework of spontaneous $T$-fluctuations both these points correspond to  $T = T_\mathrm{c}$.  Note also that Eqs.~(\ref{eq:R_EMT},\ref{eq:2D}) imply a divergence of the relative $R$-fluctuation at the superconducting transition in the form $S_\mathrm{R}/R^2\propto R^{-2}$, which is not so far from the rigorous percolation theory result~\cite{Kiss_PRL_1993} $S_\mathrm{R}/R^2\propto R^{-\lambda}$, where $\lambda=1.54\pm0.09$, that should apply in the immediate vicinity of the percolation transition.

\subsection{Why do the slow $T$-fluctuations dominate?}

Eqs.~(\ref{eq:1D},~\ref{eq:2D}) are remarkable in that the upper bound on the zero-frequency spectral density of the $R$-fluctuations is determined solely by their correlation time. Note that for the $T$-fluctuations observed in the present experiment the magnitude of the $R$-fluctuation is close to the maximum possible value. As discussed in the accompanying manuscript~\cite{Second_manuscript}, the $R$-fluctuation within a single correlation area, the size of which is given by the electron-phonon relaxation time $\tau_\text{eph}$, is not very far from the absolute upper bound given by Eq.~\eqref{eq:spectral}. Thus it is not surprising that no indication of much faster $R$-fluctuations with Ginzburg-Landau correlation time $\tau_\text{GL}$ were observed in the present experiment. 
Indeed, in 1D and 2D, we have, respectively, $S_\mathrm{R} (\omega=0)\propto\tau_\mathrm{R}^{3/2}$ and $S_\mathrm{R} (\omega=0)\propto\tau_\mathrm{R}^2$ meaning that such a slow fluctuation mechanism with $\tau_\text{eph}\gg\tau_\text{GL}$ certainly dominates the fast one at frequencies up to at least $\omega\sim\tau_\text{eph}^{-1}$. This domination is even stronger in 2D because of a stronger $1/N$ averaging. 

A more tricky question is what happens in the immediate vicinity of superconducting phase transition at $T=T_\mathrm{c}$, where the textbook scenario predicts $\tau_\text{GL}$ to diverge and thus, according to the above, the Ginzburg-Landau (GL) correlations should take over. First, it is clear that the corresponding temperature interval is extremely small, only a tiny fraction of the transition region, see Fig. 3d of the main manuscript. Second, even in the immediate vicinity of $T_\mathrm{c}$, one cannot ignore the interplay of the GL fluctuations with the $T$-fluctuations. In the present experiment, the rms $T$-fluctuation within the correlation area (given by $\tau_\text{eph}$) is on the order of 10\,mK and by far exceeds the temperature range where the relation $\tau_\text{GL}>\tau_\text{eph}$ holds. In this case, we observe that in the presence of $T$-fluctuations the system simply does not have enough time to adjust the GL fluctuations. Instead, the thermal fluctuation locally drives the system up or down in temperature across the phase transition, and, correspondingly, the GL time drastically diminishes and becomes $\tau_\text{GL}\ll\tau_\text{eph}$. This resembles a Kibble-Zurek mechanism of the transition crossing under external driving~\cite{Kibble_2007} and makes a rigorous treatment of the interplay of GL-fluctuations and $T$-fluctuations nontrivial.

\section{II. Excess noise in fluctuation region near $T_\mathrm{c}$} 

In proximity to the superconducting transition one can identify several leading corrections to the normal state conductivity originating from superconducting fluctuations, these are Maki-Thompson (MT), Aslamazov-Larkin (AL), density of states (DOS), and diffusion constant renormalization (DCR) contributions \cite{Larkinlate2005}. 
These terms are commonly derived within the linear response diagrammatic methods. The applicability of this approach relies on  the assumption of fast thermalization, $\tau_{\text{R}}(T-T_c)\ll1$, where $\tau_{\text{R}}$ is the energy relaxation time of the electrons. In this scenario, fluctuations and transport of preformed Cooper pairs occur on a background of thermal bath of quasiparticles. In the opposite regime, when the energy relaxation time of the electrons is long as compared to the time scale of the preformed pairs, $\tau_{\text{R}}(T-T_c)\gg1$, fluctuations of the conductivity adiabatically follow the fluctuations of the electron distribution function. In this section, we explore this situation and estimate the excess noise associated with long electron disequilibrium.   
In contrast to Kubo formula, an alternative approach based on the semiclassical Usadel equation formulated within the Keldysh technique of nonequilibrium superconductivity is needed for this purpose \cite{Kamenev}. This approach is especially advantageous in applications to problems with temporal and spatial fluctuations, and applications to regimes beyond the linear response such as noise \cite{kogan2008electronic}.  

\subsection*{Boltzmann-Langevin kinetic theory} 

We consider fluctuations when $T_c$ is approached from the normal state. This simplifies the theory greatly as we do not need to consider kinetics of the condensate. In the Boltzmann-Langevin scheme of kinetic equation the electron distribution function is random due to randomness of scattering. The correlation function of electron distribution fluctuations has to be established in the case-by-case situation for the relevant mechanism of relaxation under consideration. If the collision integral is taken in the relaxation-time approximation then the correlator takes a universal form \cite{kogan2008electronic}
\begin{equation}
\langle\delta f(\varepsilon,\mathbf{r},t)\delta f(\varepsilon',\mathbf{r}',t')\rangle_\Omega=\frac{4\tau_\mathrm{R}/N_0}{1+(\Omega\tau_\mathrm{R})^2}\delta(\varepsilon-\varepsilon')
\delta(\mathbf{r}-\mathbf{r}')f_\varepsilon[1-f_\varepsilon],
\end{equation}
where $N_0$ is the normal state density of states. 
As conductivity is a functional of the distribution function, $\sigma[f]$, fluctuations of $\delta f$ render corresponding fluctuations of the conductivity $\delta\sigma$. This motivates consideration of the following noise function of conductivity fluctuations    
\begin{equation}
\delta S_\sigma(\Omega)=\frac{1}{\mathcal{V}}\int \langle\delta\sigma(\mathbf{r},t)\delta\sigma(\mathbf{r}',t')\rangle e^{i\Omega(t-t')}\d(t-t')\d\mathbf{r}',
\end{equation}
where $\mathcal{V}$ is the sample volume. We use Ref. \cite{Kwak-AP2023} to establish fluctuations of conductivity in response to fluctuating electronic occupation function. It can be found in the form 
\begin{equation}
\delta\sigma(\mathbf{r},t)=\int\d\epsilon\chi(\epsilon)\delta f(\epsilon,\mathbf{r},t),\quad \chi(\epsilon)=-\frac{4e^2\mathcal{D}T^2}{\pi^2}\sum_q\int\d\omega\partial_\epsilon|C^R_{2\epsilon+\omega}(q)|^2|L^R_{\omega}(q)|^2.
\end{equation}
Here 
\begin{equation}
|C^R_{\epsilon}(q)|^2=\frac{4}{(\mathcal{D}q^2+\Gamma_\phi)^2+\epsilon^2}\,, \quad
|L^R_{\omega}(q)|^2=\frac{1}{(\mathcal{D}q^2+\tau^{-1}_\mathrm{GL}+\Gamma_\phi)^2+\omega^2}\,,
\end{equation}
describe the retarded Cooperon $C^R$, and the pair-propagator $L^R$, where we also included the dephasing rate $\Gamma_\phi$. The quantity $\chi(\epsilon)$ can be physically interpreted as a linear susceptibility of resistive fluctuations in response to fluctuations of electronic occupation. In contrast to the case of a normal metal this quantity is energy-resolved. This theory ultimately gives us an expression for the intrinsic noise that correlates fluctuations at different energies   
\begin{equation}
\delta S_\sigma(\Omega)=\frac{1}{\mathcal{V}}\left(\frac{4e^2 \mathcal{D}T^2}{\pi^2}\right)^2\frac{4\tau_\mathrm{R}/N_0}{1+(\Omega\tau_\mathrm{R})^2}
\sum_{qq'}
\iint\d\epsilon\d\omega\d\omega'
f_\epsilon[1-f_\epsilon]
\partial_\epsilon|C^R_{2\epsilon+\omega}(q)|^2
\partial_\epsilon|C^R_{2\epsilon+\omega'}(q')|^2
|L^R_{\omega}(q)|^2|L^R_{\omega'}(q')|^2
\end{equation} 
As in the case of conductivity, we are interested in the most singular term in powers of $T-T_\mathrm{c}$. Taking advantage of the separation of energy scales, $\epsilon\ll T$, it is sufficient to approximate $f_\epsilon[1-f_\epsilon]\to1$, which enables to complete $\epsilon$-integration in the closed form 
\begin{equation}
\int\d\epsilon \partial_\epsilon|C^R_{2\epsilon+\omega}(q)|^2
\partial_\epsilon|C^R_{2\epsilon+\omega'}(q')|^2=64\pi\frac{\mathcal{D}_\phi+\mathcal{D}'_\phi}{\mathcal{D}_\phi\mathcal{D}'_\phi}
\frac{3(\omega-\omega')^2-(\mathcal{D}_\phi+\mathcal{D}'_\phi)^2}{[(\mathcal{D}_\phi+\mathcal{D}'_\phi)^2+(\omega-\omega')^2]^3}
\end{equation} 
where we introduced a shorthand notations $\mathcal{D}_\phi=\mathcal{D} q^2+\Gamma_\phi$ and $\mathcal{D}'_\phi=\mathcal{D} q'^2+\Gamma_\phi$. 
The fact that noise is dominated by electrons with energies $\epsilon\ll T$ is crucial as it posteriorly justifies the applicability of the semiclassical description. 
The subsequent double-$\omega$ integrations can be also found analytically with the help of the identity 
\begin{equation}
\iint^{+\infty}_{-\infty}\frac{[3(\omega-\omega')^2-(\mathcal{D}_\phi+\mathcal{D}'_\phi)^2]\d\omega\d\omega'}
{[\mathcal{D}^2_\mathrm{GL}+\omega^2][\mathcal{D}'^2_\mathrm{GL}+\omega'^2][(\mathcal{D}_\phi+\mathcal{D}'_\phi)^2+(\omega-\omega')^2]^3}=\frac{\pi^2}
{8\mathcal{D}_\mathrm{GL}\mathcal{D}'_\mathrm{GL}(\mathcal{D}_\phi+\mathcal{D}'_\phi)(\Gamma_\mathrm{GL}-\Gamma_\phi)^3}
\end{equation}
where in analogy to above we introduced the notation $\mathcal{D}_\mathrm{GL}=\mathcal{D}q^2+\Gamma_\mathrm{GL}$. At this point we can notice that momentum integrations factorize, since the common factor $\mathcal{D}_\phi+\mathcal{D}'_\phi$ mixing momenta cancels out, thus the remaining product 
\begin{equation}
\sum_{qq'}[\mathcal{D}_\phi\mathcal{D}'_\phi\mathcal{D}_\mathrm{GL}\mathcal{D}'_\mathrm{GL}]^{-1}=\left(\sum_q[\mathcal{D}_\phi\mathcal{D}_\mathrm{GL}]^{-1}\right)^2
\end{equation}
A particular combination $\sum_q[\mathcal{D}_\phi\mathcal{D}_\mathrm{GL}]^{-1}$ appears in the expression that defines MT term \cite{Larkinlate2005}
\begin{equation}
\sigma_\mathrm{MT}=4e^2\mathcal{D}T\sum_q\frac{1}{(\mathcal{D}q^2+\tau^{-1}_\mathrm{GL})(\mathcal{D}q^2+\Gamma_\phi)}=
\frac{e^2}{8}\frac{\ln(\epsilon_\mathrm{T}/\gamma_\phi)}{\epsilon_\mathrm{T}-\gamma_\phi},\quad \epsilon_\mathrm{T}=\frac{T-T_\mathrm{c}}{T_\mathrm{c}}, \quad \gamma_\phi=\frac{\pi\Gamma_\phi}{8T_\mathrm{c}}. 
\end{equation} 
As a result, we can simply recognize that 
\begin{equation}
\delta S_\sigma(\Omega)=\frac{32\sigma^2_\mathrm{MT}\tau_\mathrm{R}/N_0 T}{\pi\mathcal{V}[1+(\Omega\tau_\mathrm{R})^2]}\left(\frac{T}{\Gamma_\mathrm{GL}-\Gamma_\phi}\right)^3.
\end{equation}
This power spectrum of conductivity fluctuations can be converted into the corresponding spectral density of current noise with the indentities $I = LE\sigma_\mathrm{MT}$ and $\delta S_\mathrm{I}=L^2E^2\delta S_\sigma$, where $E$ is the electric field and $L$ is the side length of a square sample. Then using the expression for $\sigma_\mathrm{MT}$ we arrive at the main result of this section (with restored physical units):
\begin{equation}\label{noise-MT}
\delta S_\mathrm{I}^{[\sigma]}(\Omega=0)=\frac{32}{\pi }I^2\frac{\mathcal{D}\tau_\mathrm{R}}{gL^2}
\frac{\hbar}{k_\text{B}T_\mathrm{c}}\left(\frac{k_\mathrm{B}T_\mathrm{c}\tau_\mathrm{GL}}{\hbar}\right)^3,
\end{equation}
where we introduced the dimensionless conductance of the film in the normal state $g=\hbar N_0 \mathcal{D}d$, assumed $\Gamma_\mathrm{GL}>\Gamma_\phi$, and replaced $T\to T_\mathrm{c}$ everywhere except in the Ginzburg-Landau time $\tau_\mathrm{GL}$. The corresponding AL-contribution to noise and the cross-correlated term look structurally identical to the above result.  

\subsection*{Practical estimates} 

The expression for the fluctuations of the MT-conductivity were derived based on the assumption of fluctuations of the electronic energy distribution. The analysis of experimental data suggests that the relaxation is mediated by the electron-phonon interaction, we thus identify $\tau_\mathrm{R}$ as $\tau_\mathrm{eph}$. Thus the excess noise contains three time scales: $\tau_\mathrm{GL}$, $\tau_{\phi}$ and $\tau_\mathrm{eph}$. It is of clear interest to numerically compare the respective order of magnitude of $T$-mechanism and $\sigma$-mechanism of current noise.  In the following discussion we abbreviate these terms as $\delta S^{\mathrm{[TF]}}_\mathrm{I}$ and $\delta S^{[\sigma]}_\mathrm{I}$ respectively. 

The current fluctuations mediated by $T$-mechanism at $\Omega\to0$ have the spectral density that can be calculated as follows 
\begin{equation}
\delta S^{\mathrm{[TF]}}_\mathrm{I}=E^2L^2\left(\frac{\d\sigma_\mathrm{MT}}{\d T}\right)^2\langle\delta T^2\rangle \tau_\mathrm{eph}=E^2L^2\sigma_\mathrm{MT}^2\frac{k_\mathrm{B} T^2_\mathrm{c}\tau_\mathrm{eph}}{C_\mathrm{e}L^2d(T-T_\mathrm{c})^2}=\left(\frac{8}{\pi}\right)^2 I^2\left(\frac{k_\mathrm{B}T_\mathrm{c}\tau_\mathrm{GL}}{\hbar}\right)^2\frac{k_\mathrm{B}\tau_\mathrm{eph}}{C_\mathrm{e}L^2d}.
\end{equation} 
Here we retained only the most singular term in the temperature derivative of the MT conductivity and used standard expression for the average square of temperature fluctuations that is expressed in terms of the electronic heat capacity per unit volume $C_\mathrm{e}$ \cite{LandauLifshitz1980}. One can further simplify above expression relating $C_\mathrm{e}$ with the normal state conductivity. Evaluating at $T=T_\mathrm{c}$ one finds $C_\mathrm{e}=\pi^2 k^2_\mathrm{B} N_0T_\mathrm{c}/3$ where $N_0=\sigma^{\square}_\mathrm{N}/(e^2d\mathcal{D})$. Using this relation noise formula can be reduced to:
\begin{equation}
\delta S^{\mathrm{[TF]}}_\mathrm{I}=\frac{192}{\pi^4}I^2 \frac{\mathcal{D}\tau_\mathrm{eph}}{gL^2}\frac{\hbar}{k_\mathrm{B}T_\mathrm{c}}
\left(\frac{k_\mathrm{B}T_\mathrm{c}\tau_\mathrm{GL}}{\hbar}\right)^2.
\end{equation}  
With the same conventions, using Eq. \eqref{noise-MT} we obtain for the ratio of the two noise contributions:
\begin{equation}\label{eq:ratio}
\delta S^{\mathrm{[TF]}}_\mathrm{I}/\delta S^{[\sigma]}_\mathrm{I} = \frac{6}{\pi^3}\frac{\hbar}{k_\mathrm{B}T_\mathrm{c}\tau_\mathrm{GL}}= \frac{48}{\pi^4} \frac{T-T_\mathrm{c}}{T_\mathrm{c}},
\end{equation} 
which implies that thermal fluctuations are less singular as compared to the MT-conductivity fluctuations mechanism of noise. It is important to note, that the applicability range of Eq. \eqref{noise-MT} is limited near $T_\mathrm{c}$ by the condition $\tau_\mathrm{GL} < \tau_\phi$. Thus,
although $\tau_\phi$ does not directly enter the above expressions for the noise, it sets the limit for their applicability
near $T_\mathrm{c}$. Among various possible factors (e.g. magnetic impurities), the dephasing time is intrinsically limited by the superconducting fluctuations and, according to the nonlinear fluctuations theory \cite{Larkin2001,Levchenko2010} it saturates at $T\to T_\mathrm{c}$ taking the value $\tau_\phi=(k_\text{B}T_\text{c} \sqrt{Gi}/\hbar)^{-1}$, where the Ginzburg-Levanyuk number is given by $Gi= 1/23g$. For our 5nm film of TiN with $g\approx 40$ this gives $\tau_\phi \approx 50$ ps, which is indeed much smaller time than $\tau_\mathrm{eph}$, and thus dominates the dephasing. In fact, in the present experiment, we expect the dephasing to be even stronger owing to the presence of a surface magnetic disorder~\cite{Saveskul}, that leads to $\tau_\phi \sim 5$ ps in our 5 nm film. At $\tau_\mathrm{GL} = \tau_\phi$, Eq.~\eqref{eq:ratio} predicts $\delta S^{\mathrm{[TF]}}_\mathrm{I}\sim10^{-2}\delta S^{[\sigma]}_\mathrm{I}$ for the intrinsic dephasing mechanism and $\delta S^{\mathrm{[TF]}}_\mathrm{I}\sim10^{-1}\delta S^{[\sigma]}_\mathrm{I}$ for the dephasing by magnetic impurities. This indicates that the calculated conductivity noise can be a more important fluctuation mechanism at sufficiently high temperatures. In both cases, however, this $T$ range is outside the experimentally accessed range of the steepest part of the resistive transition.

In summary, we conclude that correlational fluctuations of conductivity mediated by the fluctuating energy distribution of electrons can result in sizable and more singular at $T_\mathrm{c}$ noise contribution in comparison to the $T$-fluctuations mechanism. This conductivity noise is distinct from other intrinsic sources previously discussed in the context of superconducting transition edge sensors, see classical review in Ref. \cite{Irwin2005} and recent proposals in Ref. \cite{Wessels2021}. Note, however, that the above analysis neglects potential broadening of the resistive transition by the $T$-fluctuations, which is proposed in the main text. Possible interplay of different noise mechanisms that may arise in this case goes beyond the scope of the present work.

\vadim{
\section{III. $T$-fluctuations in the limit of $\tau_\mathrm{eph}\ll\tau_\mathrm{GL}$}
In the limit of fast e-ph relaxation the time and length scale of spontaneous $T$-fluctuations are much smaller than the GL scales $\tau_\mathrm{GL}=\pi\hbar/(8k_\mathrm{B}(T-T_\mathrm{c}))$ and $\xi_\mathrm{GL} = \sqrt{\mathcal{D}\tau_\mathrm{GL}}$. It is straightforward to estimate the RMS $T$-fluctuation at $T=T_\mathrm{c}$ in this case:
\begin{equation*}
\langle \delta T^2\rangle_\mathrm{av} = \int_0^{1/\tau_\mathrm{GL}} S_\mathrm{T}df\approx S_\mathrm{T}(f=0)\tau_\mathrm{GL}^{-1}, 
\end{equation*} 
where $S_\mathrm{T}$ is the frequency dependent spectral density of the $T$-fluctuation $S_\mathrm{T} = S_\mathrm{T}(f=0)/(1+\omega^2\tau_\mathrm{eph}^2)$. Note that the upper limit of this integral takes into account temporal averaging of the fluctuations at frequencies higher than the inverse GL-time. The correlation time of the $T$-fluctuations is given by $\tau_\mathrm{eph}$, hence:
\begin{equation*}
S_\mathrm{T}(f=0) = 4\tau_\mathrm{eph}\langle \delta T^2\rangle = 4\frac{k_\mathrm{B}T_\mathrm{c}^2}{C_\mathrm{e}\mathcal{V}_\mathrm{GL}}\tau_\mathrm{eph},
\end{equation*}
where $\mathcal{V}_\mathrm{GL}$ is the GL correlation volume. Finally:
\begin{equation}
\langle \delta T^2\rangle_\mathrm{av} = 4\frac{\tau_\mathrm{eph}}{\tau_\mathrm{GL}}\frac{k_\mathrm{B}T_\mathrm{c}^2}{C_\mathrm{e}\mathcal{V}_\mathrm{GL}}, 
\end{equation}  
}

\vadim{Next we compare the RMS $T$-fluctuation $\delta T_\mathrm{av} = \sqrt{\langle \delta T^2\rangle_\mathrm{av}}$ with the detuning from the transition point $T-T_\mathrm{c}$.}
\\

\vadim{\bf{2D case ($L,w\gg \xi_\mathrm{GL}$)}}. \vadim{In 2D: $\mathcal{V}_\mathrm{GL} = d\xi_\mathrm{GL}^2 = d\mathcal{D}\tau_\mathrm{GL}$. For an estimate we neglect the renormalization of the heat capacitance by superconducting fluctuations, $C_\mathrm{e}  = \frac{\pi^2}{3}\nu k_\mathrm{B}^2T_\mathrm{c}$, where $\nu$ is the density of states. Using the identity for the sheet conductivity $G_\square = e^2\nu d\mathcal{D}$ we obtain:
\begin{equation*}
\left(\frac{\delta T_\mathrm{av} }{T-T_\mathrm{c}}\right)^2= \frac{768}{\pi^4}\frac{k_\mathrm{B}T_\mathrm{c}\tau_\mathrm{eph}}{\hbar}\frac{e^2}{\hbar G_\square}
\end{equation*}
}

\vadim{It is convenient to express this ratio using a 2D Ginzburg-Levanyuk number $Gi^{(2)} = e^2/(23\hbar G_\square)$, so that:
\begin{equation}
\left(\frac{\delta T_\mathrm{av} }{T-T_\mathrm{c}}\right)^2 \approx 181\frac{k_\mathrm{B}T_\mathrm{c}\tau_\mathrm{eph}}{\hbar}Gi^{(2)}\label{dt_2D}
\end{equation}  
}
\\

\vadim{\bf{1D case ($L\gg \xi_\mathrm{GL}\gg w$)}}. \vadim{In 1D $\mathcal{V}_\mathrm{GL} = dw\xi_\mathrm{GL} = dw(\mathcal{D}\tau_\mathrm{GL})^{1/2}$ and the Ginzburg-Levanyuk number is  $Gi^{(1)} = \left(e^2\xi(0)/7.35wG_\square\hbar\right)^{2/3}$, where $\xi(0) = \sqrt{\pi\hbar\mathcal{D}/8k_\mathrm{B}T_\mathrm{c}}$. A straighforward derivation gives:
\begin{equation}
\left(\frac{\delta T_\mathrm{av} }{T-T_\mathrm{c}}\right)^2 \approx 58\frac{k_\mathrm{B}T_\mathrm{c}\tau_\mathrm{eph}}{\hbar}Gi^{(1)}\sqrt{\frac{Gi^{(1)}}{\epsilon}},\text{ where }\epsilon = \frac{T-T_\mathrm{c}}{T_\mathrm{c}}\label{dt_1D}
\end{equation}}

\vadim{Expressions (\ref{dt_2D}) and (\ref{dt_1D}) give an upper bound on $\tau_\mathrm{eph}$ that allows to neglect the impact of the $T$-fluctuations in 2D and in 1D (beyond the critical region, i.e. for  $\epsilon>Gi^{(1)}$). We find that $\delta T_\mathrm{av}<(T-T_\mathrm{c})$ is fulfilled for $\tau_\mathrm{eph}\lesssim 0.01 \hbar/(k_\mathrm{B}T_\mathrm{c}Gi)$, which explains the inequality used in the main text. Note that at the boundary of the critical region ($\epsilon = Gi$) this condition is more strict than  somewhat indefinite $\tau_\mathrm{eph}\ll \tau_\mathrm{GL}$.}

\section{IV. Experimental details}

\subsection{Sample parameters}
The samples are made from epitaxial TiN films exhibiting monocrystalline quality and structural uniformity. The 5-nm, 9-nm, and 20-nm thick TiN are grown on c-cut sapphire substrates at a temperature of 800$^{\circ}$C by dc reactive magnetron sputtering from a 99.999\% pure Ti target. Films are grown in an argon-nitrogen environment at a pressure of 5 mTorr and an Ar:N$_2$ flow ratio of 2:8\,sccm. Parameters of samples investigated with the noise measurements are listed in \autoref{T1}. Sample labeling preserved as in accompanying manuscript~\cite{Second_manuscript}.

\begin{table}[h!]
\caption {\label{T1} Parameters of the studied TiN samples.}
\begin{tabular}{c c c c c c c c}
\hline
\hline
    & $d$ & $L$    & $w$    & $N_\mathrm{s}$ & $R_\mathrm{n}$    & $T_\mathrm{c}$ \\ \hline
    & nm  & $\mu$m & $\mu$m &          & $\Omega$ & K     \\ \hline
A1  & 5   & 9.663  & 0.45   & 21.47    & 2.2k     & 4.073  \\
A2  &    & 2.8    & 0.11   & 25.46    & 2.3k     & 3.9   \\
A3  &    & 50     & 0.37   & 129.6    & 10k      & 4     \\
B10 & 9    & 8      & 0.064  & 125      & 1.11k    & 5.064  \\
C1 & 20   & 65  &0.28  &  240   & 800 & 5.25   \\
\hline
\hline
\end{tabular}
\end{table}

\subsection{Experimental setup for noise thermometry}
\begin{figure}[h!]
    \includegraphics[scale=1.5]{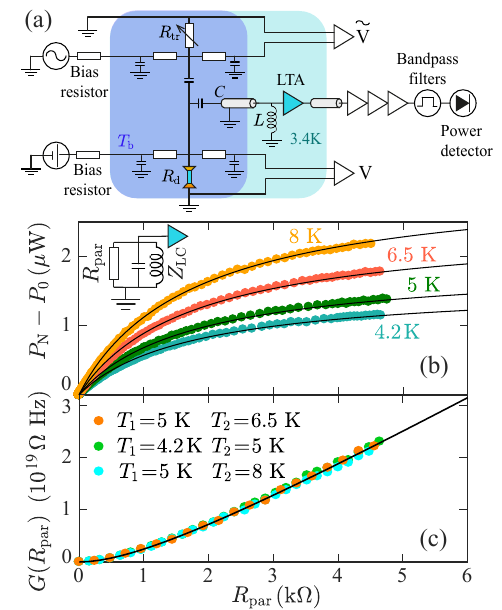}
    \caption{\label{figure_sn_1}
     Noise thermometry. (a) Schematics of the noise measurements with the resonance tank circuit and its equivalent scheme in the inset of panel (b). (b) Main: The signal from 5-nm thick sample (A3) impinging on the power detector as a function of $R_\mathrm{par}$ measured at different bath temperatures. (c) The effective total gain $G(R_\mathrm{par})$ of the experimental setup. Three sets of points correspond to three different combinations of $T_\mathrm{b}$.  Black solid lines represent the theoretical fit using the formulas \eqref{eq1}-\eqref{eq2}.}
\end{figure}

The samples are measured in a dry dilution refrigerator with a base temperature varied from 3.5 to 10\,K and in a $^3$He cryogenic insert with base temperature varied from 0.5 to 5\,K. The bath temperature $T_\mathrm{b}$ is controlled with a heater and measured with a calibrated RuO thermometer placed close to the sample. During measurements, the operating temperature is stabilized for 4 hours to achieve an accuracy of $\pm$2.5\,mK. The sample is connected through 1 nF capacitor to a high-electron-mobility transistor HEMT Avago ATF-35143, used as a variable resistor. The low-frequency transport measurements are performed in a quasi-four-terminal (q4t) circuit using stainless steel coaxial lines and low-pass resistor-capacitor filters. The low-frequency ac measurements are applied with the standard lock-in technique with a bias current in the range of 1-100\,nA and at an oscillating frequency of 8-11\,Hz. The dc transport measurements are performed in the current mode using a voltage source (Yokogawa GS200) in series with a bias resistor (see~\autoref{figure_sn_1}(a)). The ac/dc voltage is amplified by a voltage preamplifier (SR560) and measured with a voltmeter (Keysight 34461A). 

The schematic of a resonance tank circuit for the noise thermometry is shown in \autoref{figure_sn_1}(a). The output noise signal from the sample is amplified by a bespoke low-noise amplifier (LTA) and a cascade of low-noise room temperature amplifiers, then it is passed through bandpass filters and finally measured with a power detector (ZX47-60LN+). The LTA, settled at the cryostat stage with a fixed bath temperature of 3.4\,K, is based on a HEMT that converts the input current fluctuations to voltage fluctuations in a 50-$\Omega$ line. The high-frequency channel in the setup is separated from dc part by 1\,nF capacitor. The capacitance of the rf coaxial cable between the sample and the LTA, $C$ = 84.4\,pF, and the inductance of the coil, $L$ = 3\,$\mu$H, produce the resonance tank circuit. Impedance matching the sample with the LTA is carried out at the center frequency of 10\,MHz (full width at half maximum is about 1\,MHz). The measurements performed at 10\,MHz allow us to avoid the unwanted contribution of $1/f$-noise and mechanical vibrations. 

The inset in \autoref{figure_sn_1}(b) shows the equivalent circuit of the RF part of the setup. Here $R_\mathrm{par}=(R_\mathrm{res}^{-1}+R_\mathrm{tr}^{-1}+R_\mathrm{d}^{-1})^{-1}$ is the parallel resistance of the equivalent circuit, $R_\mathrm{res}=11$\,$k\Omega$ marks resistors in parallel, $R_\mathrm{tr}$ marks the resistance of the HEMT, and $R_\mathrm{d}=\dd V/\dd I$ marks the current-dependent differential resistance of the sample. The voltage noise at the input of the LTA is given by the formula~\cite{blanter2000shot} $S_\mathrm{I}^\mathrm{tot}/\abs{R_\mathrm{par}^{-1}+Z_\mathrm{LC}^{-1}}^2$, where $S_\mathrm{I}^\mathrm{tot}$ is the total current noise of the equivalent circuit and $Z_\mathrm{LC}$ is the complex impedance of the coil and cable. In general, the power spectrum $P(f)$ recorded by spectrum analyzer with a resolution bandwidth $\Delta f$ at the output of amplifiers is given by:
\begin{equation}
\label{eq1}
P(R_\mathrm{par},f) = P_\mathrm{0}(f) + G(R_\mathrm{par},f)\times S_\mathrm{I}^{tot}(R_\mathrm{par},f),
\end{equation}
where $P_\mathrm{0}(f)$ is the background power spectrum noise of the circuit, $G(R_\mathrm{par},f)$ is the effective gain including the frequency-dependent impedance of the circuit.

In the noise thermometry setup, the power noise is measured by the power detector, therefore the total noise power, the background noise power, and the effective gain are defined as $P_\mathrm{N}=(\Delta f)^{-1}\int_\mathrm{0}^\infty P(R_\mathrm{par},f) \dd f$, $P_\mathrm{0}=(\Delta f)^{-1}\int_\mathrm{0}^\infty P_\mathrm{0}(f) \dd f$, and $G(R_\mathrm{par})=(\Delta f)^{-1}\int_\mathrm{0}^\infty G(R_\mathrm{par},f) \dd f$. To calibrate the circuit we exploit the equilibrium Johnson-Nyquist (JN) noise. At a zero bias current regime, the total current noise is defined as $S_\mathrm{I}^\mathrm{tot}=4k_\mathrm{B}T_\mathrm{b}/R_\mathrm{par}+S_\mathrm{I}^\mathrm{LTA}$, where $4k_\mathrm{B}T_\mathrm{b}/R_\mathrm{par}$ is the equilibrium JN noise. Since $S_\mathrm{I}^\mathrm{tot}$ does not depend on frequency in the low-frequency limit ($f \ll k_\mathrm{B}T/2\pi\hbar$), the effective gain of the circuit can be determined as:
\begin{equation}
\label{eq2}
G(R_\mathrm{par})=\int_\mathrm{0}^\infty \frac{1}{Z_\mathrm{0}}\frac{G_\mathrm{amp}(f) Tr(f)}{\abs{R_\mathrm{par}^{-1}+Z_\mathrm{LC}^{-1}}^2} \dd f,
\end{equation}
where $Z_\mathrm{0}$ is the input impedance of the power detector, and $Tr(f)$ is the transmission function of the bandpass filters, $G_\mathrm{amp}(f)$ is the total gain of the LTA ($\sim$6\,dB), the room-temperature amplifiers ($\sim$75\,dB), and the active narrow band filter ($\sim$6\,dB). According to~Eq.~\eqref{eq2}, when $R_\mathrm{par}\rightarrow 0$ the total gain becomes zero and one can find the background noise as $P_\mathrm{0}=P_\mathrm{N}$. We can determine $G(R_\mathrm{par})$ and $S_\mathrm{I}^\mathrm{LTA}$ calibrating the circuit with the equilibrium JN noise. In the calibration procedure $R_\mathrm{par}$ is varied by applying the gate voltage on the HEMT, i.e., changing $R_\mathrm{tr}$. \autoref{figure_sn_1}(b) demonstrates the measured noise power with the subtracted background noise $P_\mathrm{N}-P_\mathrm{0}$ at different bath temperatures $T_\mathrm{b}$ (shown with symbols) and at different values of $R_\mathrm{par}$. $G(R_\mathrm{par})$ that is presented in \autoref{figure_sn_1}(c) is obtained from the fitting of $P_\mathrm{N}-P_\mathrm{0}$ at different $T_\mathrm{b}$ using Eqs.~\eqref{eq1}-\eqref{eq2}. The experimental value of $S_\mathrm{I}^\mathrm{LTA}$ is below $4\times 10^{-27}$\,A$^2$/Hz, which is significantly smaller than equilibrium JN noise, expected for the studied samples.

We measure noise power $P_\mathrm{N}^*(I,R_\mathrm{par})$ of the sample in a current-biased regime. When the current is applied to the sample, the crossover from the equilibrium JN noise, $4k_\mathrm{B}T_\mathrm{b}/R_\mathrm{d}(I=0)$, to non-equilibrium shot noise, $S_\mathrm{I}$, takes place. The transistor is pinched off ($R_\mathrm{tr}>$100 M$\Omega$) during these measurements and, hence, the total current noise and the parallel resistance are defined as $S_\mathrm{I}^\mathrm{tot}=S_\mathrm{I}+4k_\mathrm{B}T_\mathrm{b}/R_\mathrm{res}+S_\mathrm{I}^\mathrm{LTA}$ and $R_\mathrm{par}=(R_\mathrm{res}^{-1}+R_\mathrm{d}^{-1})^{-1}$. One can find the non-equilibrium shot noise of the sample using the calibration described above as follows:
\begin{equation}
S_\mathrm{I}=\frac{P_\mathrm{N}^*(I,R_\mathrm{par})-P_\mathrm{N}(I=0,R_\mathrm{par})}{G(R_\mathrm{par})}+\frac{4k_\mathrm{B}T_\mathrm{b}}{R_\mathrm{d}(I=0)}.
\label{eq2a}
\end{equation}
Note that we substitute the value of the equilibrium JN noise in Eq.~\eqref{eq2a} for both the case of the normal state and the case of the RT region. The validity of this substitution in the latter case is confirmed by measurements of $S_\mathrm{I}$ at zero bias current. Varying $T_\mathrm{b}$ at the vicinity of $T_\mathrm{c}$, we can determine $P_0$ from the $P_N(T_\mathrm{b})$ measurements taking into account that $P_0=P_\mathrm{N}$ when the sample resistance equals zero. Thus, one can find the equilibrium current noise as:
\begin{equation}
S_\mathrm{I}=\frac{P_\mathrm{N}(I=0,R_\mathrm{par})-P_0}{G(R_\mathrm{par})}-\frac{4k_\mathrm{B}T_\mathrm{b}}{R_\mathrm{res}}-S_\mathrm{I}^\mathrm{LTA}.
\end{equation}
\autoref{figure_sn_2} demonstrates that the measured voltage noise ($S_\mathrm{V}=S_\mathrm{I} R_\mathrm{d}^2$) at $I=0$ can be fully described by the expression for the equilibrium JN noise. This is the basis for substituting the JN noise into Eq.~\eqref{eq2a}.

\subsection{Results of the noise thermometry far above $T_\mathrm{c}$}

 Using the noise thermometry, we measure the current-noise spectral density $S_\mathrm{I}$, which is used to determine the spectral density of the voltage noise $S_\mathrm{V} = S_\mathrm{I} R_\mathrm{d}^2$ and the noise temperature $T_\mathrm{N} = S_\mathrm{I} R_\mathrm{d}/4k_\mathrm{B}$, where $R_\mathrm{d}$ is the current-dependent differential resistance of the sample. In the normal state $T_\mathrm{N}$ reflects the length-averaged electronic temperature $T_\mathrm{e}$. Contribution to heat relaxation due to electron diffusion into contacts can be neglected if the length of the samples $L$ is much longer than the e-ph relaxation length $L_\mathrm{eph}$, and, hence, $T_\mathrm{e}$ is considered to be uniform along $L$.

\begin{figure}[h!]
\includegraphics[scale=1]{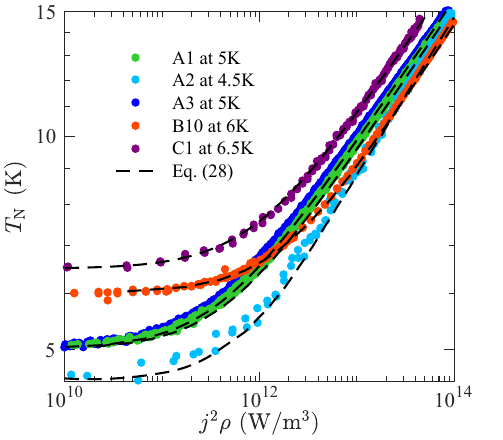}
    \caption{Noise thermometry in epitaxial TiN wires listed in \autoref{T2}. The measured noise temperature $T_\mathrm{N}$ is presented as a function of joule power density $j^2\rho$ on a log-log scale. The black dashed lines represent the fits, obtained using Eq.~\eqref{eq_ap1}. \label{figure_sn_2}}
\end{figure}

\begin{table}[h!]
\caption {\label{T2}Characteristics of thermal transport in TiN. }
\begin{tabular}{ccccccc} %
\hline
\hline
 & type  & $d$ & $\Sigma_\mathrm{eph}$  &$L_\mathrm{eph}(T_\mathrm{c})$ & $\mathcal{D}$ & $\tau_\mathrm{eph}(T_\mathrm{c})$ \\
 & & nm  & WK$^{-5}$m$^{-3}$ & $\mu$m & cm$^2$/s & ns\\ \hline
A1& bridge & 5  & 1.1$\times$10$^8$ &1.13 & 2.5 & 5.1\\
A2& bridge & 5 & 1$\times$10$^8$ & 1.35 & 2.5  & 7.3\\
A3& meander & 5  & 1.06$\times$10$^8$& 1.40  & 2.5 & 7.8\\
B10& bridge &  9 & 1.8$\times$10$^8$ & 1.5& 8.5 & 2.6\\ 
C1& meander &  20  & 0.7$\times$10$^8$& 2.5& 10 & 6.8\\ \hline
\hline
\end{tabular}
\end{table}

The results of the noise thermometry at $T>T_\mathrm{c}$ provide information about intrinsic relaxation in TiN and can be described by the standard picture of thermal relaxation in a diffusive metal wire on an insulating substrate~\cite{Huard2007}. We estimate the e-ph cooling rate $\Sigma_\mathrm{eph}$ from the experimental dependencies of the noise temperature on joule power $T_\mathrm{N}(j^2\rho)$, where $j^2\rho=IV/\mathcal{V}$ is the joule power $IV$ per unit volume $\mathcal{V} = dwL$. \autoref{figure_sn_2} shows the measured $T_\mathrm{N}$ as a function of $j^2\rho$ on a log-log scale for samples A1-A3, B10, and C1. The plateau in the $T_\mathrm{N}(j^2\rho)$-dependence corresponds to the small heating regime, when $T_\mathrm{N}$ remains close to the bath temperature $T_\mathrm{b}$. Here, $T_\mathrm{b}$ is set slightly higher than $T_\mathrm{c}$ to keep the superconducting TiN samples in the normal state.
In the intense heating regime, when $T_\mathrm{N}\gg T_\mathrm{b}$, we observe similar power-law dependencies for all samples with $j^2\rho \propto T_\mathrm{N}^5$. To describe the experimental dependencies more accurately we use the steady-state heat balance equation, considering $T_\mathrm{N}\approx T_\mathrm{e}$:
\begin{equation}
\label{eq_ap1}
\dv{x}\left(\kappa_\mathrm{e}\dv{T_\mathrm{e}}{x} \right)=-j^2\rho+ \Sigma_\mathrm{eph}\left( T_\mathrm{e}^5-T_\mathrm{ph}^5\right) 
\end{equation}
where $T_\mathrm{e}$ and $T_\mathrm{ph}$ are the electron and phonon temperatures in the metal bridge, $\kappa_\mathrm{e}$ is the electronic thermal conductivity given by the Wiedemann-Franz law. In this analysis, we assume that $T_\mathrm{ph} = T_\mathrm{b}$, since the estimated Kapitza resistance effect is negligible for thin TiN films on the sapphire substrate (see for details Section V in Supplemental Material). Eq.~\eqref{eq_ap1} provides the fits for the experimental data presented in \autoref{figure_sn_2} with the fitting parameter $\Sigma_\mathrm{eph}$. Taking into account the diffusion coefficient $\mathcal{D}=8k_\mathrm{B}T_\mathrm{c}\xi(0)^2/\pi\hbar$ and $\Sigma_\mathrm{eph}$ we estimate the electron-phonon relaxation length $L_\mathrm{eph} = \sqrt{\mathcal{L}/\rho_\mathrm{n}5\Sigma_\mathrm{eph}T_\mathrm{c}^{3}}$ and the e-ph relaxation time $\tau_\mathrm{eph}=L_\mathrm{eph}^2/\mathcal{D}$ at $T_\mathrm{c}$. The experimental parameters, $\Sigma_\mathrm{eph}$, $\mathcal{D}$, $L_\mathrm{eph}$ at $T_\mathrm{c}$, and $\tau_\mathrm{eph}$ at $T_\mathrm{c}$ are listed in \autoref{T2}.~These estimations are in agreement with previously observed energy relaxation time measurements in disordered TiN films at $T_\mathrm{c}$~\cite{Kardakova2013}, and they reveal relatively slow electron-phonon relaxation in TiN. Extrapolated values of $\tau_\mathrm{eph}$ to room temperature (10 - 40\,fs) are also close to the values of $\tau_\mathrm{eph}$ obtained in similar TiN films via the pump-probe technique \cite{Diroll2020}.

\subsection{Results of the noise thermometry along the R(T)-dependence}

\begin{figure}[h!]
\includegraphics[scale=1.5]{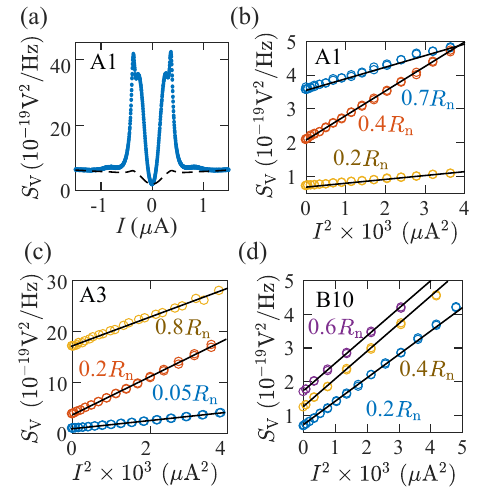}
    \caption{Giant noise at the resistive transition in samples A1 (a-b), A3 (c), and B10 (d). (a) The dependence of the voltage noise $S_\mathrm{V}$ on the bias current $I$ at $R(T)=0.4R_\mathrm{n}$ in comparison to the Johnson-Nyquist equilibrium noise, shown by the dashed line. (b-d) The voltage noise $S_\mathrm{V}$ vs. squared current $I^2$ measured at the resistive transition for a few values of $T_\mathrm{b}$. \label{figure_sn_3}}
\end{figure}

A completely different picture of the non-equilibrium noise is observed at the resistive transition.~\autoref{figure_sn_3}(a) represents a typical giant noise signal as a function of bias current $I$. At intermediate currents, the observed noise (symbols) is much higher than the equilibrium Johnson-Nyquist noise with $T_\mathrm{N} = T_\mathrm{b}$, the latter is shown by the dashed line.~\autoref{figure_sn_3}(b-d) shows the spectral density of the voltage noise $S_\mathrm{V}$ as a function of $I^2$ measured for samples A1, A3, and B10. Here, $S_\mathrm{V}(I)$, measured at different values of $T_\mathrm{b}$ at which $R < R_\mathrm{n}$, scales with the current as $S_\mathrm{V}\propto I^2$ (see the black lines in~\autoref{figure_sn_3}). These quadratic dependencies are observed close to a linear response transport regime, with the differential resistance increasing by less than 20\% with $I$. Thus the giant voltage noise at the resistive transition originates from resistance fluctuations in equilibrium~\cite{kogan2008electronic}, for which $S_\mathrm{V} = I^2 S_\mathrm{R}$, where $S_\mathrm{R}$ is the spectral density of the resistance noise, the so-called modulation noise. 

\par
\begin{figure}[h!]
\includegraphics[scale=1]{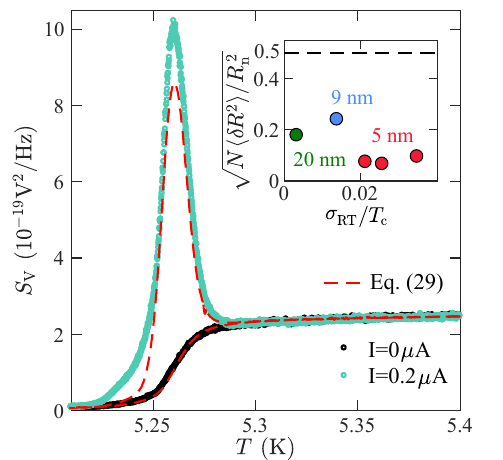}
    \caption{\label{figure_sn_4}
    A comparison to the model of spontaneous $T$-fluctuations. Main: Temperature dependencies of $S_\mathrm{V}$ \elmira{measured at resonance frequency 10\,MHz} at a zero bias current (black symbols) and at $I$ = 0.2\,$\mu$A (blue symbols). The data is presented for sample 20-nm thick sample (C1). The red dashed line is a fit with Eq.~\eqref{eq_Sv_20nm}. Inset: relative fluctuation of the resistance in a correlation volume as a function of the width of the resistive transition $\sigma_\mathrm{RT}$ defined as a $T$-range between $0.2 R_\mathrm{n}$ and $0.8 R_\mathrm{n}$ and normalized to $T_\mathrm{c}$. The inset illustrates the idea that the narrower the resistive transition, the greater $R$-fluctuation in the correlation volume.}
\end{figure}

 To describe the giant noise along the $R(T)$-dependence we elaborate the model of temperature fluctuations ($T$-fluctuations), which implies a one-to-one thermodynamic correspondence between $R$-fluctuations and $T$-fluctuations. Here, we use the empirical expression $\langle \delta T ^2\rangle=\abs{ \dd R/\dd T}^{-2}\langle \delta R ^2\rangle$~\cite{Voss1976,Kogan1985} describing the noise due to $T$-fluctuations in superconducting bolometers~\cite{Ekstrom1995, Hoevers2000}. The variance of an electronic temperature is well-known in thermodynamics, and it can be expressed as $\langle \delta T^2\rangle = k_\mathrm{B}T_\mathrm{e}^2/C_\mathrm{e} \mathcal{V} $~\cite{LandauLifshitz1980}, where $C_\mathrm{e}$ is the electronic specific heat capacity. The variance of $R$-fluctuations can be defined using the correlation time as $\langle\delta R^2\rangle = \int_\mathrm{0}^\infty S_\mathrm{R}(f) \dd f = \int_\mathrm{0}^\infty S_\mathrm{R}(0) \dd f/(1+(2\pi f\tau_\mathrm{R})^2) = S_\mathrm{R}(0)/4\tau_\mathrm{R}$. If $T$-fluctuations give rise to $R$-fluctuations, we can detect them by measuring the voltage noise $S_\mathrm{V}$ at frequencies $f\ll 1/2\pi\tau_\mathrm{R}$. To make these fluctuations visible in noise measurements, a finite bias current $I$ is applied to a sample, since $\delta S_\mathrm{V}(0) = I^2 S_\mathrm{R}(0)$, and the voltage noise is given by:
\begin{equation}
S_\mathrm{V} = 4 k_\mathrm{B} T_\mathrm{e} R_\mathrm{d}+ 4k_\mathrm{B}\frac{\tau_\mathrm{R}}{C_\mathrm{e}}I^2\abs{\dd R/\dd T_\mathrm{e}}^2T_\mathrm{e}^2.
\label{eq_Sv_20nm}
\end{equation}
Here, the first and the second terms correspond to the equilibrium Johnson–Nyquist noise and the excess noise at RT, respectively.~\autoref{figure_sn_4} shows the temperature dependencies of $S_\mathrm{V}$ at two different bias currents: $I$ = 0\,$\mu$A (black symbols) and $I$ = 0.2\,$\mu$A (blue symbols). The data is presented for the 20-nm thick TiN sample (C1). At a zero bias current, $S_\mathrm{V}$ coincides with the Johnson–Nyquist expression $(4 k_\mathrm{B} T_\mathrm{e} R_\mathrm{d})$. At a finite bias current, $S_\mathrm{V}$ exceeds the Johnson–Nyquist noise peaking at a maximum of $\dd R(T)/\dd T$. Such temperature dependence of $S_\mathrm{V}$ agrees well with the previously reported data~\cite{Voss1980,Knoedler1983,Zhang2008}. To estimate the magnitude of voltage fluctuations, we assume that the current-biased superconductor near the resistive transition remains in local thermal equilibrium. In other words, we assume that Joule heating slightly raises $T_\mathrm{e}$ above the bath temperature, i.e. $\delta T_\mathrm{e} = T_\mathrm{e} - T\ll T$, and the current–voltage dependence of the device simply follows the temperature dependence of the resistance $R(T\approx T_\mathrm{e})$. Taking into account $C_\mathrm{e} = \pi^2 k_\mathrm{B}^2T_\mathrm{e} N_\mathrm{0}/3$ for the free-electron gas~\cite{Kittel}, where the estimated DOS at Fermi level $N_\mathrm{0}\simeq$ 90\,eV$^{-1}$nm$^{-3}$, we fit the data in~\autoref{figure_sn_4} with Eq.~\eqref{eq_Sv_20nm} and obtain results remarkably close to the experiment (see the red dashed line). 

What about the observed magnitude of $R$-fluctuations in each correlation volume? The instantaneous value $R(t)$ at RT obeys the condition $0\leq R(t)\leq R_\mathrm{n}$, where $R_\mathrm{n}$ is the normal-state resistance. For a given average resistance $\langle R\rangle$, the variance is bounded by $\langle \delta R^2\rangle\leq \langle R\rangle\left(R_\mathrm{n} - \langle R\rangle\right)$. The equality takes place only for bimodal fluctuations between $R = 0$ and $R = R_\mathrm{n}$, the so-called case of the telegraph noise~\cite{kogan2008electronic}. Thus, the maximum variance of the resistance within a single correlation volume corresponds to $\langle\delta R_i^2\rangle = (R_i^n)^2/4$, where $R_i^n=R_\mathrm{n}/N$ is the normal state resistance of the $i$-th correlation volume. For simplicity, we assume a 1D case with $N$ correlation volumes in the sample, in which the fluctuations occur independently. Here $N = L/L_{c}$, and $L_{c} = L_\mathrm{eph}$ for $T$-fluctuations. Now, it is instructive to compare the expected variance with the measured $R$-fluctuations. The total fluctuation in the sample is given by $\langle \delta R^2\rangle = N \langle\delta R_i^2\rangle$, which predicts the maximum value of $\sqrt{N \langle \delta R^2\rangle/R_\mathrm{n}^2} = 1/2$ for the telegraph noise. Using the empirical value of $\langle \delta R^2\rangle$, we plot $\sqrt{N \langle\delta R^2 \rangle/R_\mathrm{n}^2}$ as a function of the RT width, $\sigma_\mathrm{RT}$ (see the inset of~\autoref{figure_sn_4}). It is interesting to note that the narrower the transition, the greater the fluctuation, and it is quite close to the absolute maximum in 9-nm (B10) and 20-nm (C1) thick samples. In the main text we argue that although $T$-fluctuations disappear in the large sample limit (in thermodynamic sense), they can be considered as a fundamental lower bound on the width of RT.

\subsection{Experimental setup for noise spectroscopy}

To get insights into the dynamics of $R$-fluctuations, we perform the spectroscopic measurements of the giant noise at the resistive transition. For this purpose we extend the frequency range of measurements removing the coil at the input of the LTA and the bandpass filters (see~\autoref{figure_sn_1}(a)). The noise power spectrum $P(R_\mathrm{par},f)$ measured via the spectrum analyzer (R$\&$S FSP or Agilent E4446A) within the 100-MHz bandwidth is described by Eq.~\eqref{eq1}. Since $R_\mathrm{d}$ depends on current and temperature at resistive transition, the tricky part for such measurements is matching the sample to $50\,\Omega$-line. To minimize the impedance mismatch, we anchor the value of $R_\mathrm{par}$ by tuning the HEMT resistance at each operating point. Thus, the fixing of $R_\mathrm{par}$ value enables $ G(R_\mathrm{par},f)$ to be insensitive to a change in $R_\mathrm{d}$.

\begin{figure}[h!]
    \includegraphics[scale=1.25]{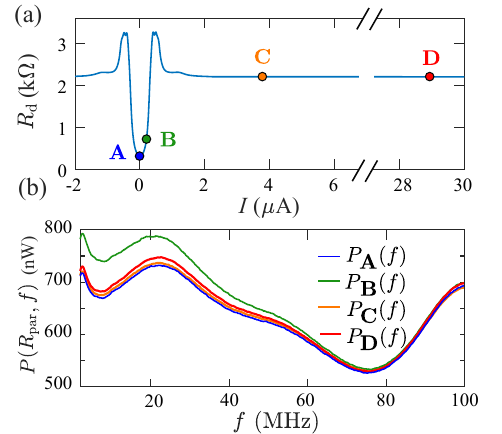}
    \caption{\label{figure_sn_5}
    Noise spectroscopy. (a) The current dependence of the differential resistance $R_\mathrm{d}(I)$ of sample A1 (the blue line). Here, the symbols show the different operating regimes: (\textbf{A}) the superconducting state at $I=0$, (\textbf{B}) the superconducting state at $I\neq 0$, (\textbf{C}) the small heating regime in the normal state, (\textbf{D}) the intense heating regime in the normal state. (b) The noise power spectrum measured by the spectrum analyzer at different operating points at $R_\mathrm{par}=50\,\Omega$.}
\end{figure}

We define four operating regimes for measuring $P(R_\mathrm{par},f)$ from the current dependence of $R_\mathrm{d}$. \autoref{figure_sn_5}(a) represents $R_\mathrm{d}(I)$ for sample A1 at $T_\mathrm{b}=4$\,K, which is obtained by the differentiation of the current–voltage curve. The symbols show the following regimes: (\textbf{A}) the superconducting state at $I = 0$, (\textbf{B}) the resistive state at $I = 0.1$\,$\mu$A, (\textbf{C}) the normal state at the small heating ($I = 4$\,$\mu$A), (\textbf{D}) the normal state at the intense heating ($I = 30$\,$\mu$A). In mode \textbf{A}, all finite resistances of the circuit produce the equilibrium Johnson-Nyquist noise, which does not depend on frequency up to $f \approx k_\mathrm{B} T_\mathrm{b}/h$ ($\simeq$0.1\,THz at $T_\mathrm{b}=4$\,K). In the normal state (\textbf{C} and \textbf{D}), the bias current $I$ causes an increase in $T_\mathrm{e}$, as shown before with the noise thermometry measurements, and the noise spectrum is also assumed to be white here. Meanwhile, in the resistive state (the mode \textbf{B}) the bias current $I$ induces the giant noise.

\par \autoref{figure_sn_5}(b) demonstrates the noise power spectra measured by the spectrum analyzer at different operating modes. One can see that $P_\mathrm{\textbf{A}}(f)$ and $P_\mathrm{\textbf{C}}(f)$ are indistinguishable and therefore can be interchangeable in our case. To perform the calibration procedure, we define $ G(R_\mathrm{par},f)$ at fixed $R_\mathrm{par}$, subtracting two white spectra with different magnitudes obtained in the normal state. In this case, according to Eq.~\eqref{eq1} the gain of the circuit is defined as $G(R_\mathrm{par},f) = (P_\mathrm{\textbf{D}}(f)-P_\mathrm{\textbf{C}}(f))R_\mathrm{n}/4k_\mathrm{B}\Delta T_\mathrm{e}$. We estimate the difference in electron temperature $\Delta T_\mathrm{e}\simeq$ 25.5\,K at operating points \textbf{C} and \textbf{D} using the noise thermometry data. The giant voltage noise can be found from the subtraction of the spectra $P_\mathrm{\textbf{B}}(f)-P_\mathrm{\textbf{A}}(f)$ as:

\begin{equation}
    \delta S_\mathrm{V}(f)=\frac{P_\mathrm{\textbf{B}}(f)-P_\mathrm{\textbf{A}}(f)}{G(R_\mathrm{par},f)}R_\mathrm{d}^2.
    \label{eq_delta_sv}
\end{equation}
Note that $\delta S_\mathrm{V}(f)$ here does not include the equilibrium JN noise.

\begin{figure}
\includegraphics[scale=1.25]{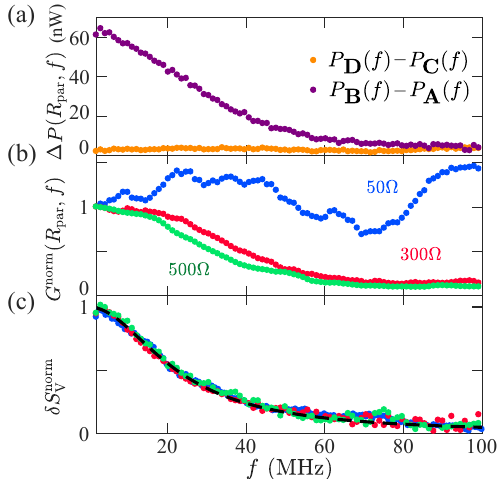}
    \caption{\label{figure_sn_6}
Noise in the TiN wire at the resistive transition and the calibration procedure. (a) The difference of the power spectra $P_\mathrm{\textbf{D}}(f)-P_\mathrm{\textbf{C}}(f)$ (orange symbols) and $P_\mathrm{\textbf{B}}(f) - P_\mathrm{\textbf{A}}(f)$ (purple symbols) measured at $R_\mathrm{par}$ = 50\,$\Omega$. The frequency dependence of $ G(R_\mathrm{par},f)$ (b) and $\delta S_\mathrm{V}$ (c) normalized to the value at zero frequency for different values of $R_\mathrm{par}$. The black line shows the fit by the Lorentzian function.}
\end{figure}

\begin{figure}
\includegraphics[scale=0.9]{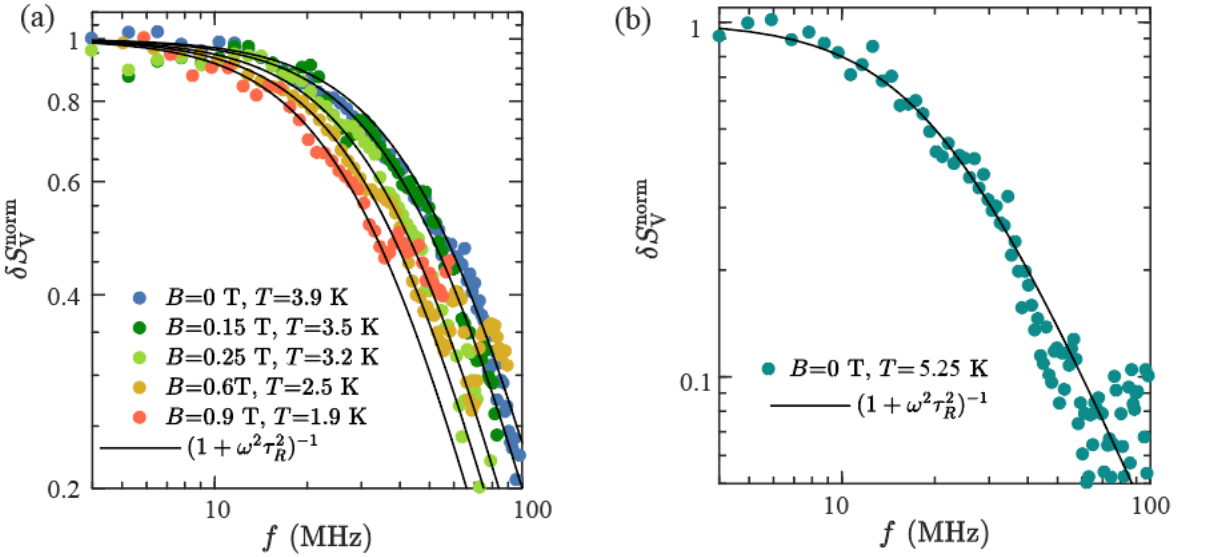}
    \caption{\label{figure_sn_7}
Normalized spectra of $\delta S_\mathrm{V}$ on a log-log scale within 4 to 100\,MHz for sample A2 (a)  and C1 (b). The black solid lines represent Lorentzian fits used to obtain $\tau_\mathrm{R}$. For sample A2, $\delta S_\mathrm{V}^\mathrm{norm}$ are measured at different temperature $T$ and value of magnetic field $B$. The T-dependence of obtained values of $\tau_\mathrm{R}$ for sample A2 are presented in Fig. 3(c) in the main text. For sample C1 the spectrum is measured at zero magnetic field. The obtained value of $\tau_\mathrm{R}$ from Lorentzian fit (black solid line) is about 8\,ns.
}
\end{figure}

\autoref{figure_sn_6}(a) demonstrates the difference in the noise power spectra $\delta P(f)$ at fixed $R_\mathrm{par}$ = 50\,$\Omega$. Here, the subtraction $P_\mathrm{\textbf{D}}(f)-P_\mathrm{\textbf{C}}(f)$ reflects the weak frequency dependence of $G(R_\mathrm{par},f)$, meanwhile $P_\mathrm{\textbf{B}}(f)-P_\mathrm{\textbf{A}}(f)$ demonstrates the strong frequency dependence of amplified $\delta S_\mathrm{V}(f)$. Next, to check the universality of $\delta S_\mathrm{V}(f)$-dependence we perform the noise spectroscopy at different $R_\mathrm{par}$ values: 50~$\Omega$, 300~$\Omega$, and 500~$\Omega$. As shown in \autoref{figure_sn_6}(b), $ G(R_\mathrm{par},f)\propto P_\mathrm{\textbf{D}}(f)-P_\mathrm{\textbf{C}}(f)$ crucially depends on the choice of $R_\mathrm{par}$. The strong frequency dependence is also observed for the subtractions $P_\mathrm{\textbf{B}}(f)-P_\mathrm{\textbf{A}}(f)$ at $R_\mathrm{par}$ = 300\,$\Omega$ and $P_\mathrm{\textbf{B}}(f)-P_\mathrm{\textbf{C}}(f)$ at $R_\mathrm{par}$ = 500\,$\Omega$ (not shown here). Nevertheless, the spectrum $\delta S_\mathrm{V}(f)$ is not affected by the choice of the value of $R_\mathrm{par}$ (see \autoref{figure_sn_6}(c)). The measured noise spectrum can be fitted by the Lorentzian function with a fitting parameter $\tau_\mathrm{R}$. The data, presented for sample A1 at $T_\mathrm{b}$ = 4\,K, are fitted by the Lorentzian function $( 1+\left(2\pi f \tau_\mathrm{R}\right)^{2})^{-1}$, where $\tau_\mathrm{R}$ = 6.5$\pm$0.5\,ns, see the dashed line fit. The spectra for other samples (Sample A2 and C1) are presented in \autoref{figure_sn_7}.

\section{V. Kapitza resistance}

To understand the influence of Kapitza resistance on the thermal transport in TiN, we estimate the thermal resistance $Z_\mathrm{K}$ for TiN-Al$_2$O$_3$ interface at $T$ = 5\,K. $Z_\mathrm{K}$ can be obtained by summing over two transverse and one longitudinal acoustic modes~\cite{Swartz1989}:

\begin{equation}
Z_\mathrm{K}=\frac{30 \hbar^3}{\pi^2 k_\mathrm{B}^4}\left(\frac{v^{-2}_\mathrm{1L}+2v^{-2}_\mathrm{1T} +v^{-2}_\mathrm{2L}+2 v^{-2}_\mathrm{2T} }{\left(v^{-2}_\mathrm{1L}+2v^{-2}_\mathrm{1T} \right) \left(v^{-2}_\mathrm{2L}+2 v^{-2}_\mathrm{2T} \right)}\right)T^{-3}
\end{equation}
where, $v_\mathrm{iL}$ and $v_\mathrm{iT}$ are the longitudinal and transverse sound velocities in $i$ medium. To calculate $Z_\mathrm{K}$ we use values of the sound speeds reported for Al$_2$O$_3$ ($v_\mathrm{1L}$ = 10.89\,km/s, $v_\mathrm{1T}$ = 6.45\,km/s~\cite{Swartz1989}). The velocities $v_\mathrm{2L}$ = 10.95\,km/s and $v_\mathrm{2T}$ = 5.6\,km/s in TiN are obtained from elastic constants~\cite{Kim1992}. We find that $Z_\mathrm{K}$ is about 2.5$\times 10^{-5}$\,W K$^{-1}$m$^{-2}$ for the thermal interface of TiN-Al$_2$O$_3$ at $T$ = 5\,K. From the experimental data the thermal resistance is given by $\dd T_N/\dd P_\mathrm{2D}$, where $P_\mathrm{2D} = IV/wL$ is the joule power per unit area. Finally, Kapitza resistance is an order of magnitude smaller than the estimated thermal resistance $\dd T/\dd P_\mathrm{2D}$, which is within 1.8-5$\times 10^{-4}$\,W K$^{-1}$m$^{-2}$ at 5\,K and we neglect it in the analysis.

\section{VI. Energy relaxation time due to electron diffusion into contacts}
The thermal transport in short metal bridges ($L\ll L_\mathrm{eph}$) can be described by the time-dependent heat-balance equation~\cite{Huard2007}:
\begin{equation}
\label{eq_time}
C_\mathrm{e}\pdv{T_\mathrm{e}}{t} = \pdv{x}\left(\kappa_\mathrm{e}\pdv{T_\mathrm{e}}{x}\right)
\end{equation}
Taking into account $ \kappa_\mathrm{e}=C_\mathrm{e}\mathcal{D}$, Eq.~\eqref{eq_time} can be transformed into:

\begin{equation}
\label{eq_time1}
 \pdv{u}{t}= \mathcal{D}\pdv[2]{u}{x},
\end{equation}
where $u(x,t) \equiv T_\mathrm{e}^2(x,t)$. The solution of the equation is usually obtained by using the separation of variables method ~\cite{tikhonov2013equations}: $u(x,t) = X(x)T(t)$, where $X(x)$ is a function of the variable $x$, and $T(t)$ is a function of the variable $t$. Substitution of $u(x,t)$ in Eq.~\eqref{eq_time1} provides $T'(t)/\mathcal{D} T(t) = X''(x)/X(x) = -\lambda$, where $\lambda$ is a constant. The substitution of the zero boundary conditions $u(0,t) = u(L,t) = 0$ into the $x$-dependent part ($X''(x)+\lambda X(x) = 0$) yields the solution $X(x) = sin(\sqrt{\lambda} x)$, where $\lambda = (\pi m /L)^2$ with $m = 1,2,3..$. This procedure allows us to find the proper relaxation time of $T$-fluctuations, associated with diffusion into contacts ($\tau_\mathcal{D}$). For $m = 1$, the slowest component of $t$-dependent part of the equation is obtained, and its solution $T(t) \propto e^{-t/\tau_\mathcal{D}}$ determines the limiting value of $\tau_\mathcal{D} = L^2/(\pi^2 \mathcal{D})$ at $T \rightarrow  0$ in our experiment.


%